\documentclass[10pt,a4paper]{article}
\pdfoutput=1
\usepackage{curves,multicol}
\usepackage[export]{adjustbox}
\usepackage{subcaption}
\usepackage[makeroom]{cancel}
\usepackage{scalerel}
\usepackage{natbib}
\usepackage{graphicx}
\usepackage{hyperref}
\usepackage{authblk}
\usepackage{vmargin}
\usepackage{amssymb}
\usepackage{amsfonts}
\usepackage{multirow}
\usepackage{framed}
\usepackage{latexsym}
\usepackage{array}
\usepackage{tensor}
\usepackage{url}
\usepackage[dvipsnames]{xcolor}
\usepackage{graphicx}
\usepackage{epstopdf, epsfig}
\usepackage{booktabs}
\usepackage{commath}
\usepackage{rotating}
\usepackage{tabularx}
\usepackage{bibentry}
\usepackage{doi}
\usepackage{lineno}
\usepackage{inconsolata}
\usepackage{makecell}
\usepackage{etc/jlcode}
\usepackage{titlesec}

\providecommand{\keywords}[1]{\textbf{\textit{Keywords ---}} #1}
\newcommand{\fref}[1]{Fig.~\ref{#1}}
\newcommand{\tref}[1]{Tab.~\ref{#1}}

\newcommand{\sref}[1]{\S\ref{#1}}

\newcolumntype{A}{ >{$} r <{$} @{} >{${}} l <{$} } % Align table cells for equations
\newcolumntype{L}[1]{>{\raggedright\let\newline\\\arraybackslash\hspace{0pt}}m{#1}}
\newcolumntype{C}[1]{>{\tiny\centering\let\newline\\\arraybackslash\hspace{0pt}}m{#1}}
\newcolumntype{R}[1]{>{\raggedleft\let\newline\\\arraybackslash\hspace{0pt}}m{#1}}
\newcolumntype{N}{@{}m{0pt}@{}}

\makeatletter
\newcommand{\smallsym}[2]{#1{\mathpalette\make@small@sym{#2}}}
\newcommand{\make@small@sym}[2]{%
  \vcenter{\hbox{$\m@th\downgrade@style#1#2$}}%
}
\newcommand{\downgrade@style}[1]{%
  \ifx#1\displaystyle\scriptstyle\else
    \ifx#1\textstyle\scriptstyle\else
      \scriptscriptstyle
  \fi\fi
}
\makeatother

\hypersetup{
	unicode=true,
	pdftitle={A descriptive title},
	pdfnewwindow=true,
	colorlinks=true,
	pdfborder={0 0 0},
	linkcolor=blue,
	linktoc=all,
	citecolor=black,
	urlcolor=blue,
	breaklinks=false,
}
\setmarginsrb { 0.8in}  % left margin
              {0.25in}  % top margin
              { 0.8in}  % right margin
              { 0.5in}  % bottom margin
              {  20pt}  % head height
              {0.25in}  % head sep
              {   9pt}  % foot height
              { 0.3in}  % foot sep
\graphicspath{{img/}}

\title{\textbf{WaterLily.jl: A differentiable and backend-agnostic Julia solver for incompressible viscous flow around dynamic bodies}}
\author[1]{Gabriel D. Weymouth}
\author[1,2,\footnote{\href{mailto:b.font@tudelft.nl}{\texttt{b.font@tudelft.nl}}}]{Bernat Font}

\affil[1]{Faculty of Mechanical Engineering, Delft University of Technology, Delft, Netherlands}
\affil[2]{Barcelona Supercomputing Center, Barcelona, Spain}
\titlelabel{\thetitle.\,\,}

\date{\today}

\begin{document}

{\let\newpage\relax\maketitle}
\setlength{\parindent}{0pt}
\setlength{\parskip}{8pt}

\date{\vspace{-20pt}}
\begin{abstract}
Integrating computational fluid dynamics (CFD) solvers into optimization and machine-learning frameworks is hampered by the rigidity of classic computational languages and the slow performance of more flexible high-level languages. In this work, we introduce WaterLily.jl: an open-source incompressible viscous flow solver written in the Julia language. An immersed boundary method is used to enforce the effect of solid boundaries on flow past complex geometries with arbitrary motions. The small code base is multidimensional, multiplatform and backend-agnostic, ie. it supports serial and multithreaded CPU execution, and GPUs of different vendors. Additionally, the pure-Julia implementation allows the solver to be fully differentiable using automatic differentiation. The computational cost per time step and grid point remains constant with increasing grid size on CPU backends, and we measure up to two orders of magnitude speed-up on a supercomputer GPU compared to serial CPU execution. This leads to comparable performance with low-level CFD solvers written in C and Fortran on research-scale problems, opening up exciting possible future applications on the cutting edge of machine-learning research.
\end{abstract}

\begin{small}
  \noindent
  \\
  \textbf{WaterLily.jl repository:} \href{https://github.com/WaterLily-jl/WaterLily.jl}{\texttt{https://github.com/WaterLily-jl/WaterLily.jl}}\\
  \textbf{Manuscript repository:} \href{https://github.com/WaterLily-jl/WaterLily.jl_CPC_2024}{\texttt{https://github.com/WaterLily-jl/WaterLily.jl\_CPC\_2024}}
\end{small}

\section{Introduction}
During the last decade, the computational fluid dynamics (CFD) community has embraced the surge of machine learning (ML) and the new developments in hardware architecture, such as general-purpose graphics-processing units (GPUs). Hence, classic CFD solvers based on low-level programming languages (C, Fortran) and CPU memory-distributed execution are being adapted to accommodate these new tools. On one hand, the integration of high-level ML libraries and low-level CFD solvers is not straight-forward, aka. the two-language problem \citep{Churavy2022}. When deploying an ML model online with the CFD solver, data exchange is often performed at disk level, significantly slowing down the overall runtime because of disk read and write operations. An improved way to exchange data is performed through memory, either using Unix sockets \citep{Rabault2019, Font2021}, message-passing interface (MPI) \citep{Guastoni2023}, or an in-memory distributed database \citep{Kurz2022,Font2024,Font2025}, which increases the software complexity. On the other hand, porting classic CFD solvers to GPU is also a non-trivial task which often requires the input and expertise of GPU vendors \citep{Romero2022}. Still, the CFD community has been an active asset in this transition, and it currently offers a rich variety of open-source multi-GPU solvers as summarized in \tref{tab:solvers}.

\begin{table}[!ht]
\centering
\begin{tabular}{llll}
    \hline
    \thead{Name} & \thead{Application} & \thead{Method} & \thead{Language} \\
    \hline
    \makecell{CaNS\\ \cite{Costa2018}} & \makecell{Incompressible canonical flows on\\ rectilinear grids} & FDM & Fortran/OpenACC \\\hline
    \makecell{GAL{\AE}XI\\ \cite{Kempf2024}} & Compressible flows on unstructured grids & DG & CUDA-Fortran \\\hline
    \makecell{MFC\\ \cite{Bryngelson2021}} & \makecell{Compressible multi-phase flows\\ on structured grids} & FVM & Fortran+Fypp/OpenACC \\\hline
    \makecell{nekRS\\ \cite{Fischer2022}} & Incompressible flows on unstructured grids & SEM & C++/OCCA \\\hline
    \makecell{Oceananigans.jl\\ \cite{Ramadhan2020}} & Geophysical flows & FVM & Julia (calls C++ libraries) \\\hline
    \makecell{OpenSBLI\\ \cite{Lusher2021}} & \makecell{Code-generation system for compressible\\ flows on structured grids} & FDM & \makecell{Python +\\ CUDA/OpenCL} \\\hline
    \makecell{PyFR\\ \cite{Witherden2015}} & \makecell{Compressible/incompressible flows on\\ unstructured grids} & FR & \makecell{Python + C/OpenMP, \\ CUDA, OpenCL} \\\hline
    \makecell{RHEA\\ \cite{Jofre2023}} & Compressible flows on rectilinear grids & FDM & C++/OpenACC \\\hline
    \makecell{SOD2D\\ \cite{Gasparino2024}} & \makecell{Compressible/incompressible flows on\\ unstructured grids} & SEM & Fortran/OpenACC \\\hline
    \makecell{STREAmS\\ \cite{Bernardini2021}} & \makecell{Compressible canonical wall-bounded flows\\ on rectilinear grids} & FDM & CUDA-Fortran \\\hline
\end{tabular}
\caption{Examples of multi-GPU open-source CFD solvers. Methods are abbreviated as: finite difference method (FDM), discontinuous Galerkin (DG), spectral element method (SEM), finite volume method (FVM), and flux reconstruction (FR).}\label{tab:solvers}
\end{table}

In this context, Julia \citep{Bezanson2017} emerges as an open-source programming language specifically designed for scientific computing which can help tackle such software challenges. The dynamic just-in-time compiled language allows high-level libraries and low-level code to co-exist without compromising computing performance. Moreover, its excellent meta-programming capabilities maximize code re-usability. Two pertinent packages in Julia for the present work are the automatic differentiation (AD) package \href{https://github.com/JuliaDiff/ForwardDiff.jl}{ForwardDiff.jl} and the heterogeneous kernel generation package \href{https://github.com/JuliaGPU/KernelAbstractions.jl}{KernelAbstractions.jl}, \citep{RevelsLubinPapamarkou2016,Churavy2023}. These packages enable any high-level code written in pure Julia to be differentiated automatically and executed efficiently on a wide-range of backends such as multithreaded CPU, NVIDIA, AMD, and many HPC systems \citep{Churavy2022,Churavy2024}.

In this work, we leverage this capability to develop \href{https://github.com/WaterLily-jl/WaterLily.jl}{WaterLily.jl}, a new differentiable CFD solver with heterogeneous execution. Unique among the solvers detailed in \tref{tab:solvers}, WaterLily's pure Julia implementation enables efficient performance-critical code in a compact and uniform framework. In less than 1000 lines of code, WaterLily is a fully-differentiable and easily extendable CFD solver that can run in CPU or GPU architectures of different vendors without compromising performance. The numerical methods and software design of WaterLily are reported in sections \sref{sec:numerical_methods} and \sref{sec:software_design}. The performance of the solver is analysed in \sref{sec:performance} with benchmarks and kernel profiling. Validation on canonical cases is performed in \sref{sec:validation}, and different test cases showcasing notable features of the solver are shown in \sref{sec:applications}. Finally, conclusions and expectations are presented in \sref{sec:conclusions}.

\section{Numerical methods}\label{sec:numerical_methods}
WaterLily uses the boundary data immersion method (BDIM) to simulate the fluid flow around immersed bodies \citep{Weymouth2011,Maertens2014,Lauber2022}. The preceding references give the precise mathematical formulation, as well as detailed validation of the immersed-boundary method's accuracy. To summarize the approach, the momentum equation defined over the fluid domain
\begin{equation}
    \mathcal{\dot F}:\ \dot u_i = -p_{,i} - (u_i u_j)_{,j}+\nu u_{i,jj} \quad \forall i,j \in 1\ldots n
\end{equation}
is integrated in time and convolved with a prescribed body velocity defined over the solid domain
\begin{equation}
    \mathcal{B}:\ u_i = V_i \quad \forall i \in 1\ldots n
\end{equation}
resulting in a single meta-equation valid over the whole space. In these equations, $u_i$ are the velocity components in a $n$-dimensional flow, $p$ is the pressure scaled by the fluid density, $\nu$ is the fluid viscosity, $V_i$ is the body velocity, indices after commas indicate spatial derivatives, and summation is used over repeated indices. As with these equations, WaterLily can be applied to simulations of any number of dimensions $n$, although we typically restrict applications to 2D and 3D flows.

The immersed-boundary thickness $\epsilon$ defines the region directly affected by the prescribed body velocities, but the flow inside this region still obeys the fluid dynamic equations with second-order accuracy \citep{Maertens2014}. The transition between these two regions is defined by the properties of the immersed surface, specifically the signed-distance $d$ and normal $\hat n$ from any point in space to the closest surface point. This, along with the body velocity $V_i$, defines the local meta-equation.

WaterLily implements the governing equation using a finite-volume approach on a uniform Cartesian grid with staggered velocity-pressure variable placement. Since all grid cells are identical, no grid information is stored. Second-order central differences are used for the pressure and diffusion terms, while a flux-limited Quick scheme is used on the convective term. While explicit turbulence models have been used for specific projects, the core WaterLily package is model-free, making it an implicit Large Eddy Simulation (iLES) solver \citep{Margolin2006}.

Finally, the momentum equation is integrated in time using an explicit predictor-corrector update scheme \citep{Lauber2022}. The velocity is restricted to be incompressible (divergence-free, $u_{i,i}=0$) using a pressure projection scheme at each step. The resulting Poisson equation has spatially varying coefficients in the presence of immersed boundaries, and it is solved using a geometric multi-grid method \citep{Weymouth2022}. The time step is adapted automatically to ensure a stable Courant–-Friedrichs–-Lewy (CFL) number.

\section{Software design}\label{sec:software_design}

Julia’s flexible and fast programming capabilities enabled the implementation of WaterLily to have many special features in a minimal codebase. The most important Julia features for implementing the solver are (i) the dynamic typing and just-in-time compilation, (ii) the meta-programming capabilities, and (iii) the rich open-source packages which are based on these features.

Dynamic type dispatching enables simple functions (such as broadcasting or reductions operations on arrays) to be written at a high-level by the user, while intermediate Julia libraries, and ultimately the compiler, will specialize the code for efficient execution on a particular architecture (CPU or GPU). This is the basis for the forward-mode AD package in Julia - which recompiles AD-unaware code using a general Dual number type to generate efficient high-order functions for the \textit{exact} derivative and extends automatically to nested derivatives. We use AD extensively \textit{within} WaterLily to define all the properties of the immersed geometry from a user-defined signed-distance function and a coordinates-mapping function. Moreover, the solver is itself differentiable, enabling exact derivatives with respect to simulation inputs to be efficiently generated. Having access to the exact derivatives using AD is especially important when dealing with nested functions, where finite difference method would likely fail, such as when using data-driven methods to optimize solver coefficients as demonstrated in \cite{Weymouth2022}.

For more specialized tasks, WaterLily uses Julia's meta-programming features to generate code that produces an individual specific kernel. The kernel can be used to offload the computing workload into a GPU, or to run it in a multithreaded CPU environment depending on the available system architecture. As an example, the gradient of the \jlinl{n}-dimensional pressure field \jlinl{p} is applied to the velocity field \jlinl{u} as follows

\begin{minipage}{\linewidth}
\begin{jllisting}
for i in 1:n  # apply pressure gradient
    @loop u[I, i] -= c[I, i] * (p[I] - p[I - ∂(i)]) over I in inside(p)
end
\end{jllisting}
\end{minipage}

where \jlinl{∂(i)} is a function defining a Cartesian index step in the direction \jlinl{i}, \jlinl{c} are the coefficients in the pressure-Poisson matrix arising from the discretization scheme, and \jlinl{inside(p)} provides the range of Cartesian indices \jlinl{I} in the pressure field to loop over (excluding ghost cells). For example, if \jlinl{size(p) == (10, 10)}, then \jlinl{inside(p)} yields a range of \jlinl{CartesianIndices((2:9, 2:9))}. When applying the \jlinl{@loop} macro to this expression, the following kernel is produced based on the KernelAbstractions.jl (KA) package API \citep{Churavy2023}

\begin{minipage}{\linewidth}
\begin{jllisting}
@kernel function kern_(u, c, p, i, @Const(I0)) # automatically generated kernel
    I = @index(Global, Cartesian)
    I += I0
    @fastmath @inbounds u[I, i] -= c[I, i] * (p[I] - p[I - ∂(i)])
end
\end{jllisting}
\end{minipage}

which is subsequently launched with the auto-generated call

\begin{minipage}{\linewidth}
\begin{jllisting}
kern_(get_backend(u))(u, c, p, i, inside(p)[1] - oneunit(inside(p)[1]), ndrange=size(inside(p)))
\end{jllisting}
\end{minipage}

Note that \jlinl{@kernel}, \jlinl{@index}, \jlinl{@Const} and \jlinl{get_backend} are part of the KA API and ultimately generate the appropriate kernel based on the backend inferred by \jlinl{get_backend(u)}. KA can specialize generally written kernels to specific hardware architectures using the following Julia packages and related backends: CUDA.jl and GPUArrays.jl \citep{Besard2018,Besard2019} for CUDA kernels (NVIDIA GPUs), AMDGPU.jl \citep{Samaroo2025} for ROCm kernels (AMD GPUs), oneAPI.jl \citep{Besard2022opeAPI} for oneAPI kernels (Intel GPUs), and Metal.jl \citep{Besard2022Metal} for Metal kernels (Apple GPUs). Moreover, generally written KA kernels can also run on CPU including multithreading support.

WaterLily kernels use a Cartesian-index based parallelization across the global memory, and the \jlinl{@Const(I0)} argument passes the ghost-cell offset information into the kernel. The workgroup size for the parallelization of the range of Cartesian indices (\jlinl{ndrange}) is automatically inferred based on the size of each dimension in \jlinl{ndrange}. Moreover, the backend of the working arrays, such as \jlinl{u} or \jlinl{p}, is specified by the user through the \jlinl{mem} (for memory) keyword argument when creating a \jlinl{Simulation} object. Hence, with a simple flag, the CFD simulation can be run on a CPU or a GPU from different vendors. Currently, WaterLily has been successfully tested on both NVIDIA and AMD GPUs. Similarly, the precision of the simulation is specified with the keyword argument \jlinl{T}, which for example can be set to \jlinl{Float32} (single) or \jlinl{Float64} (double) precision.

The highest-level data-type in WaterLily is the \jlinl{Simulation} type, which holds information about the fluid through the \jlinl{Flow} type, and the immersed body (or bodies) through the \jlinl{AutoBody} type. Hence, to set up a simulation, the user must specify the size of the Cartesian grid as well as other optional properties such as characteristic length and velocity, fluid viscosity, and type of boundary conditions (slip by default, otherwise a convective outlet or a periodic condition can be selected too). On the other hand, the \jlinl{AutoBody} type holds the signed-distance function as well as the coordinates mapping for moving boundaries. More detailed examples on how to set up a simulation are available in \sref{sec:applications}.

\section{Performance analysis} \label{sec:performance}
The performance of the solver is assessed on three different cases with increasing level of complexity: (1) the Taylor--Green vortex (TGV) at $Re=1600$, (2) flow past a fixed sphere at $Re=3700$, (3) and flow past a moving circular cylinder at $Re=1000$. Note that the cases range from a flow free of solid boundaries to a flow containing a dynamic body.

\begin{enumerate}
\item The TGV case consists of a developing flow transitioning to turbulence in a $L^3$ triple-periodic cubic domain. The initial condition for the TGV velocity vector field $\vec{u}_0$ is prescribed as
\begin{align}
u_0 &= -U\sin(\kappa x)\cos(\kappa y)\cos(\kappa z) \\
v_0 &= U\cos(\kappa x)\sin(\kappa y)\cos(\kappa z) \\
w_0 &= 0,
\end{align}
where $U=1$ is the characteristic velocity, and $\kappa=2\pi/L$ is the wavenumber. As in \cite{Dairay2017}, our computational domain is the half-domain and we apply symmetry boundary conditions to lower the cost of the simulation. To test different grid resolutions, we select $L=\left\{2^6,2^7,2^8,2^9\right\}$ resulting into grids of 0.26, 2.10, 16.78, and 134.22 million of degrees of freedom (DOF), ie. grid cells, respectively.
\item The static sphere case features a sphere in uniform flow with diameter systematically increased using $L=\left\{2^3,2^4,2^5,2^6\right\}$ cells to benchmark the scaling performance. The computational domain size is set relative to the diameter size, using relative dimensions $16L\times6L\times6L$, resulting in benchmark simulations with 0.29, 2.36, 18.87, and 150.99 million DOF. Slip conditions are applied on the lateral boundaries and a convective outlet is applied on the downstream plane.
\item For the dynamic circular cylinder case, the diameter is increased using $L=\left\{2^4,2^5,2^6,2^7\right\}$ cells in a $9L\times6L\times2L$ domain, resulting in 0.44, 3.54, 28.31, 226.49 million DOF.
The transverse and downstream conditions are the same as the sphere, but a periodic boundary condition is applied on the spanwise direction of the cylinder.
\end{enumerate}
Benchmarks and profiling are both conducted using Julia 1.11, WaterLily 1.4, and 32-bits (single) floating-point precision (FP32).

\subsection{Benchmarks}
\begin{figure}[!t]
  \centering
  \begin{subfigure}[t]{0.32\linewidth}
      \centering
      \includegraphics[width=\linewidth]{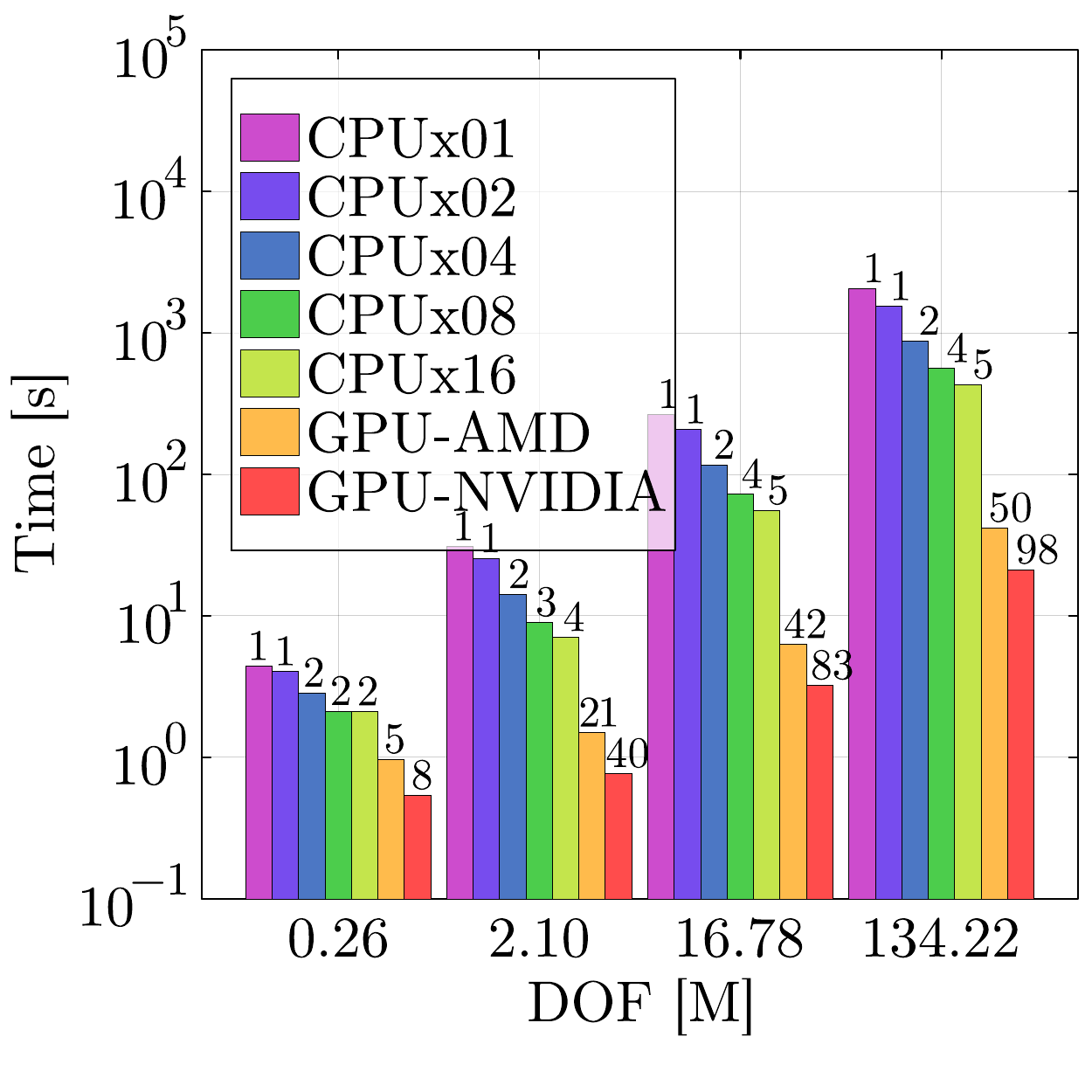}
  \end{subfigure}
  \begin{subfigure}[t]{0.32\linewidth}
    \centering\hspace*{-0.2cm}
    \includegraphics[width=\linewidth]{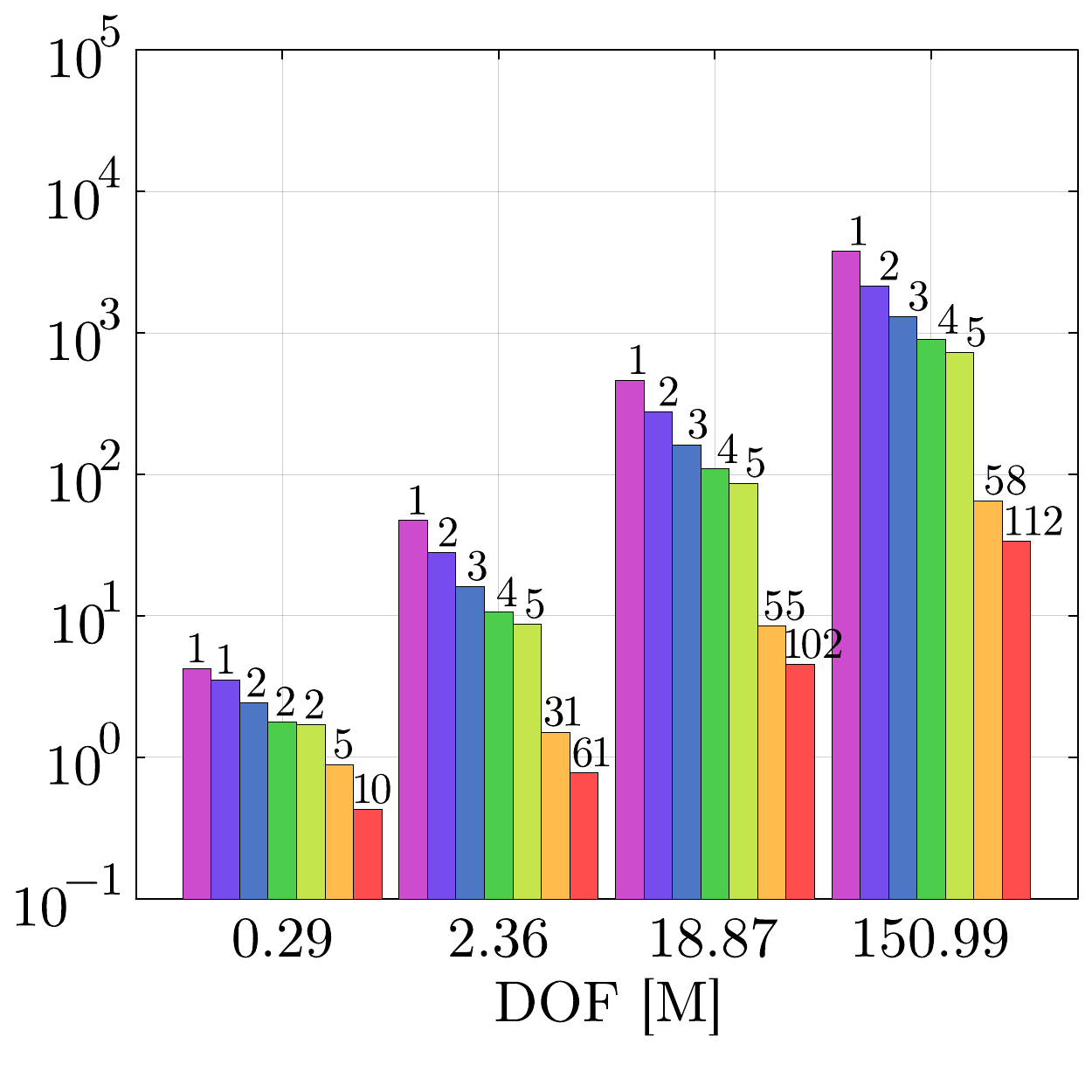}
  \end{subfigure}
  \begin{subfigure}[t]{0.32\linewidth}
    \centering\hspace*{-0.2cm}
    \includegraphics[width=\linewidth]{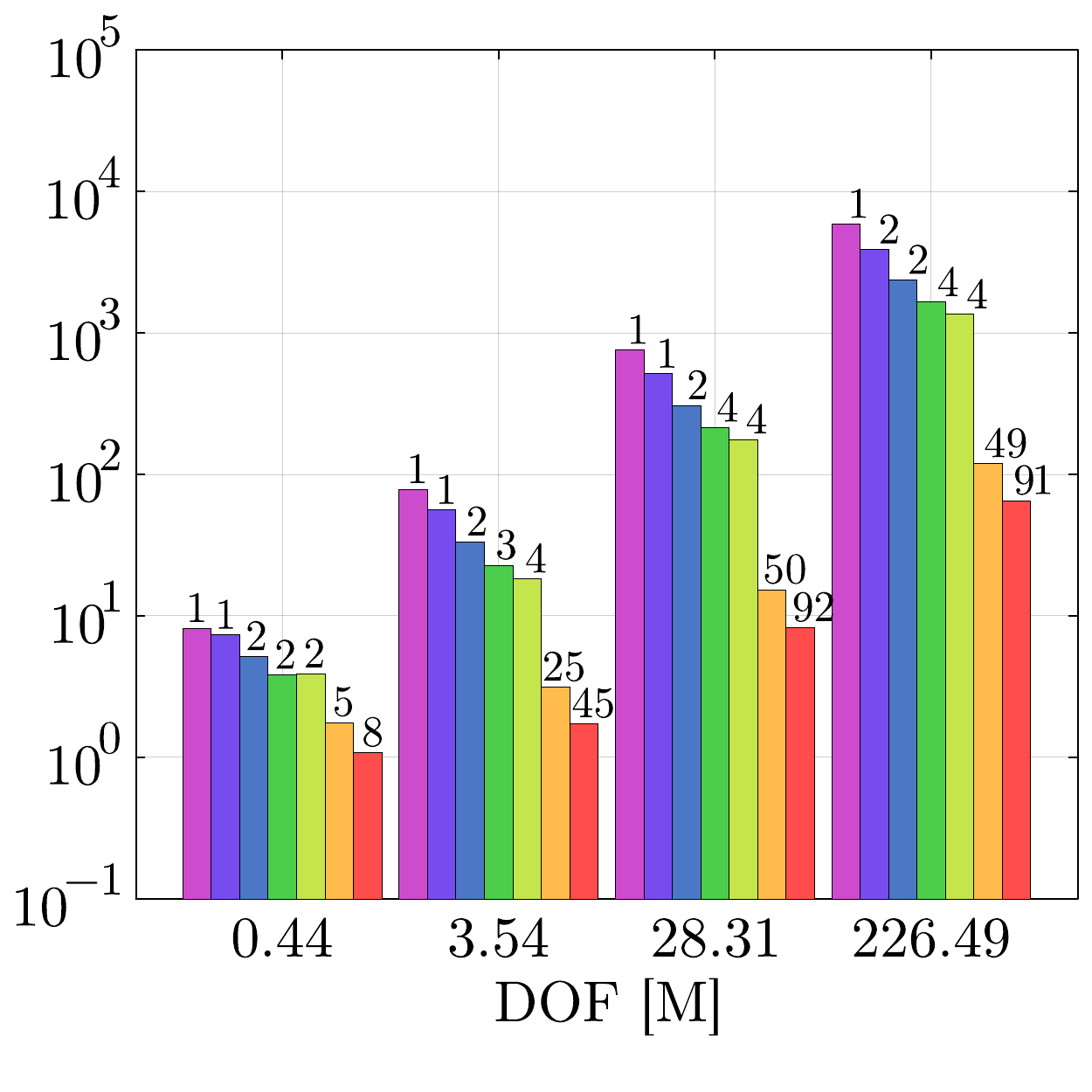}
  \end{subfigure}
  \begin{subfigure}[t]{0.32\linewidth}
    \centering
    \includegraphics[width=\linewidth]{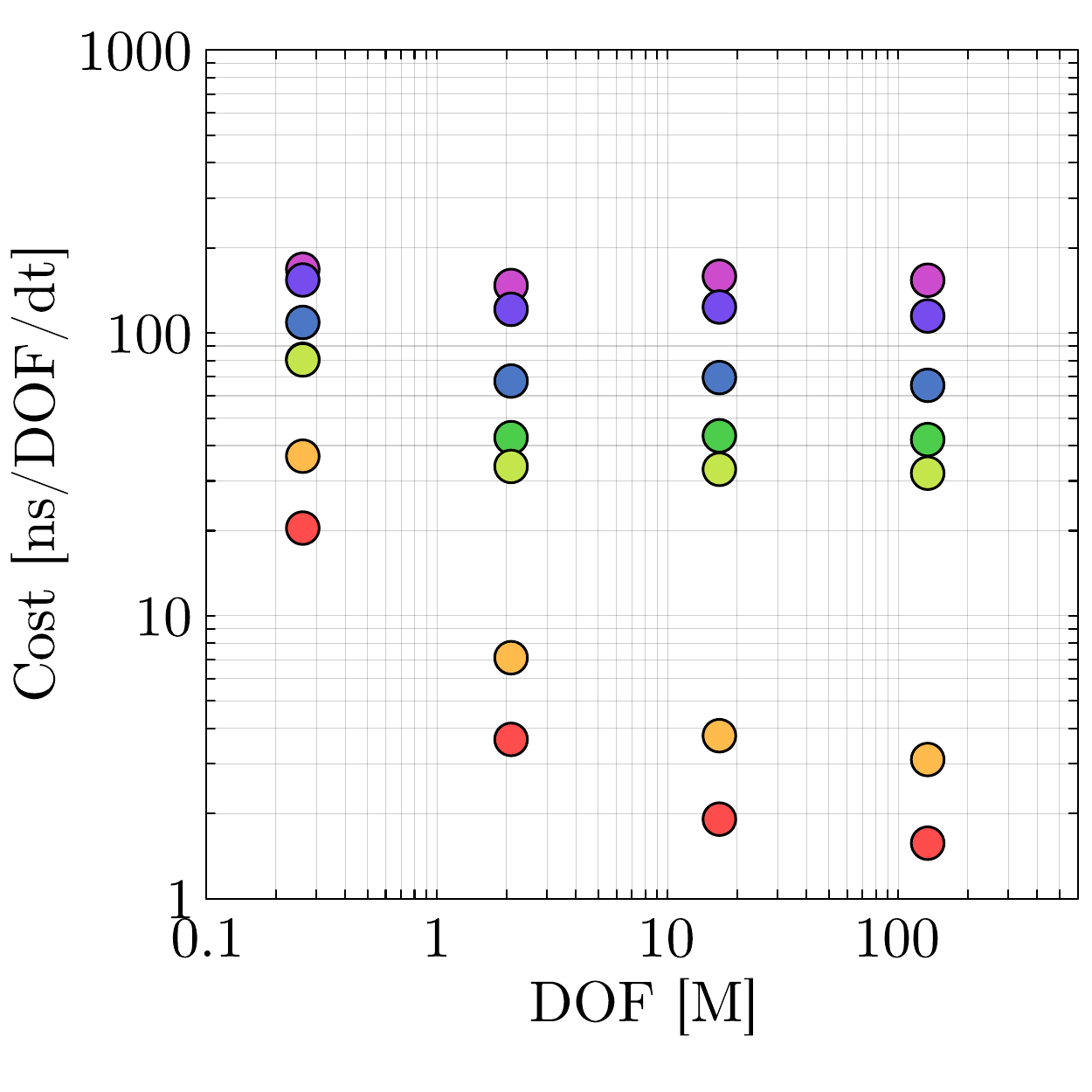}
    \caption{TGV\hspace*{-1em}}
  \end{subfigure}
  \begin{subfigure}[t]{0.32\linewidth}
    \centering\hspace*{-0.2cm}
    \includegraphics[width=\linewidth]{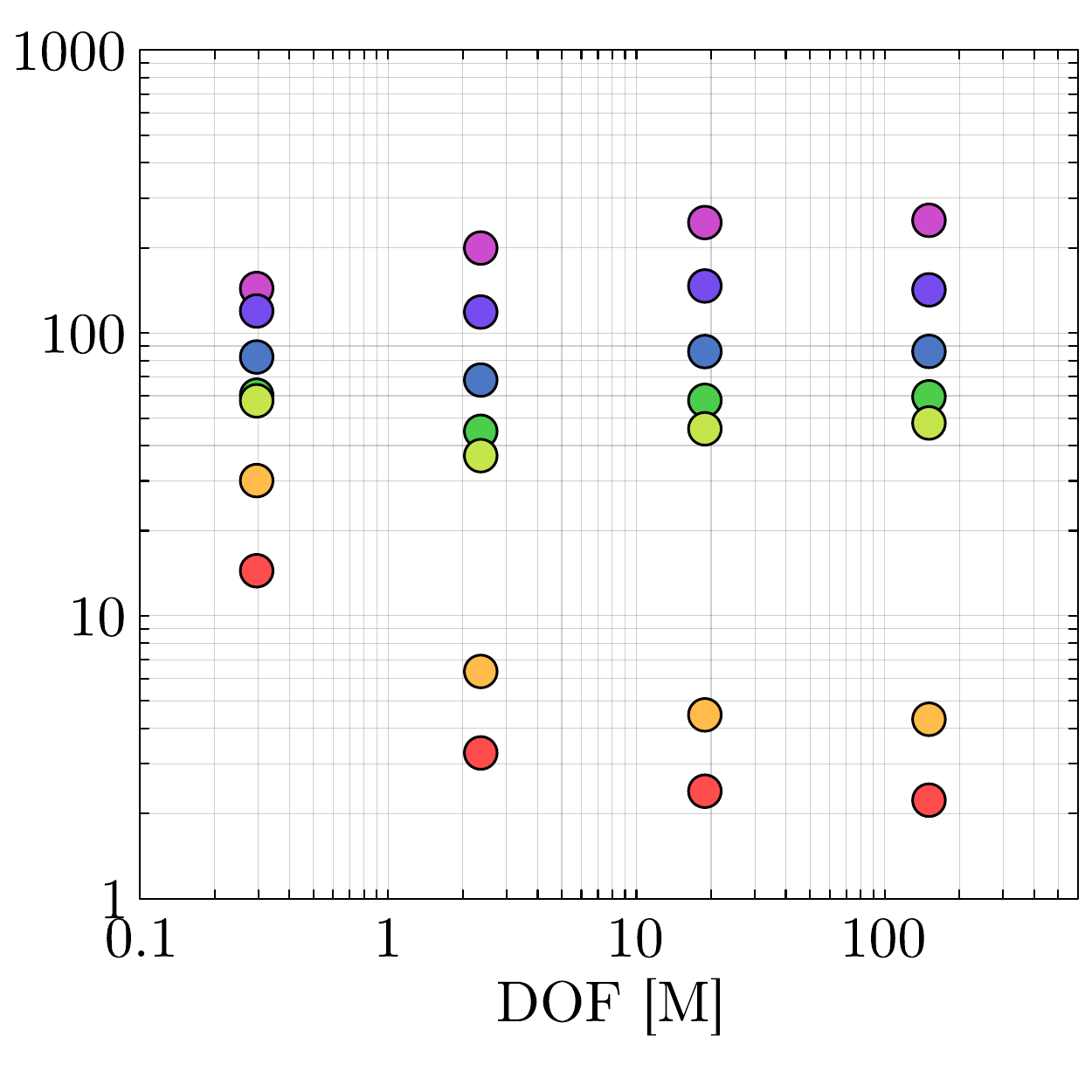}
    \caption{Sphere\hspace*{-1em}}
  \end{subfigure}
  \begin{subfigure}[t]{0.32\linewidth}
    \centering\hspace*{-0.2cm}
    \includegraphics[width=\linewidth]{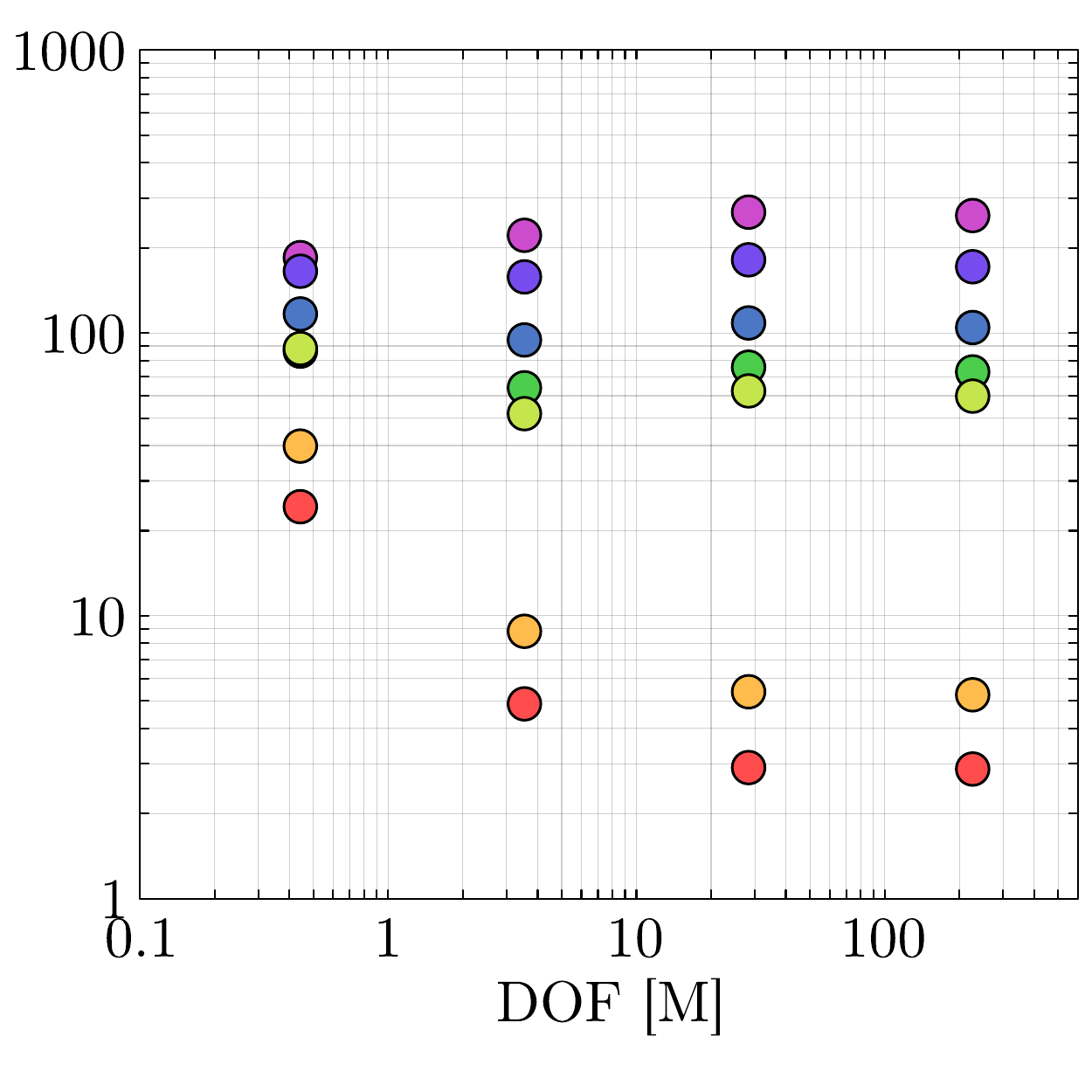}
    \caption{Moving cylinder\hspace*{-1em}}
  \end{subfigure}
	\caption{\textbf{Top}: Time to run 100 time steps in single precision for the different cases and grid sizes. The CPU execution comprises multiple number of threads from single thread (serial) to 16 threads (multithreading). The speed-up for each case with respect to the serial execution, time(CPUx01)/time(X), is shown above each bar. \textbf{Bottom}: Cost, defined as the execution time per grid DOF and time step, on the different cases and grid levels.}
	\label{fig:benchmarks}
\end{figure}

The benchmark of the three cases is measured by timing the execution of 100 time steps using different backends, as displayed in \fref{fig:benchmarks}. Benchmarks have been run on (i) an accelerated node of the Marenostrum5 supercomputer including Intel Xeon Platinum 8460Y @ 2.3GHz cores (CPU backend) and an NVIDIA Hopper H100 GPU (GPU-NVIDIA backend), and (ii) an AMD MI250X GPU of the LUMI supercomputer (GPU-AMD backend). In terms of FP32 peak performance, the NVIDIA GPU offers 60 TFLOPs (ie. $60\cdot10^{12}$ floating-point operations per second) and the AMD GPU offers 50 TFLOPs, approximately. Both GPUs have 64GB of vRAM.

On the CPU backend, increasing the number of threads enables faster simulations in a linear trend that slowly stagnates to a factor of 5x approximately. A possible reason behind the low multithreading scalability is that KA is designed for GPU kernels, and CPU kernels include additional statements to translate GPU-focused semantics and make them compatible with CPU backends, which can negatively impact multithreading scalability. In addition, note that, except for the 1st-level grids, the CPU backend does not yield a larger speed-up when increasing the grid size.

In contrast, the speed-up of the GPU with respect to the serial CPU execution is greatly increased as the GPU vRAM is filled, approximately reaching two orders of magnitude in speed-up factor. The effect of improved performance when maximizing the GPU memory load is also observed in other CFD codes \citep{Kempf2024,Gasparino2024}. Analysing the compute and memory workloads of the dominant kernels of the cylinder case (more details in the profiling section), we observe that the memory throughput approximately doubles from the 1st to the 3rd-level grid. Since the solver is memory-bounded, the increase in memory bandwidth improves the overall performance. The cost, ie. the execution time per grid point and time step, clearly demonstrates this behaviour on the GPU backends. On the other hand, the CPU backends show an approximately constant cost for grids larger that 1M DOF, being this the desired behaviour.

In terms of single-GPU cost performance, the TGV case yields 1.57 ns/DOF/dt on the largest grid using the NVIDIA H100 GPU, which is competitive with other CFD solvers such as SOD2D \citep{Gasparino2024} or NekRS \citep{Fischer2022}. It can also be observed that the performance on the NVIDIA GPU is approximately two times better than on the AMD GPU. While the specifications on memory bandwidth and peak compute performance of both GPUs are similar, we observe that CUDA kernels are faster compared to the ROCm ones. This can be either attributed to the performance of the kernel code automatically generated by KA or, at a deeper level, to a more efficient utilization of compute resources by CUDA compared to ROCm. The AMD GPU is designed to operate at a thermal design power (TDP) of 400W, and an average consumption of 380W is measured on the largest grid. On the other hand, the NVIDIA GPU has a TDP of 700W, and an average consumption of 420W is measured. Taking into account the execution time on each GPU, the total energy consumption on the NVIDIA GPU is lower than on the AMD GPU.

\subsection{Profiling}
\begin{figure}[!t]
  \centering
  \begin{subfigure}[t]{0.32\linewidth}
      \centering
      \includegraphics[width=\linewidth]{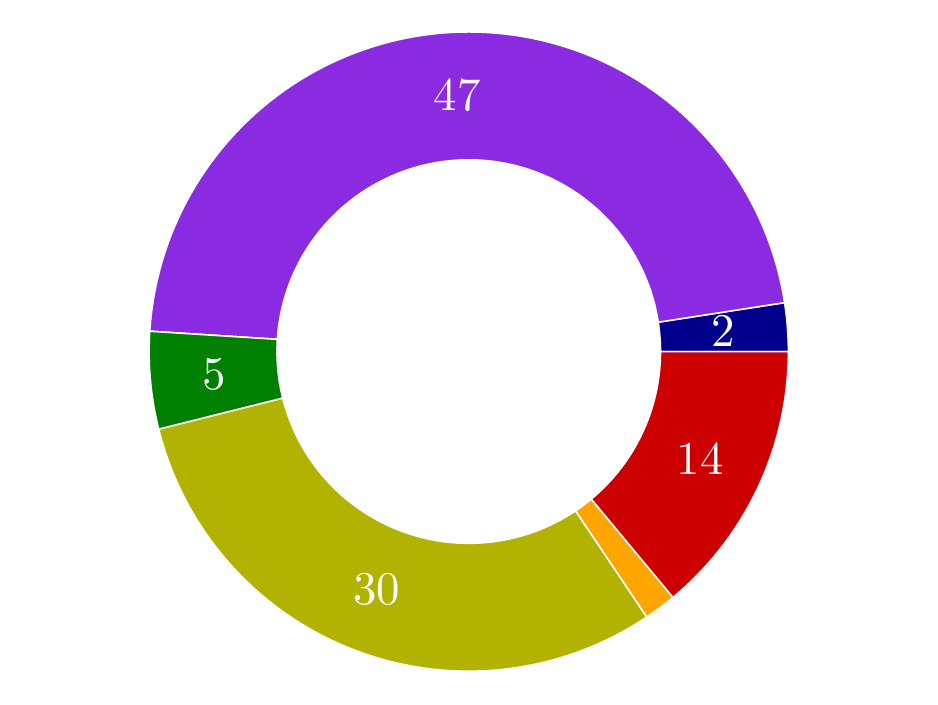}
      \caption{TGV\hspace*{1em}}
  \end{subfigure}
  \begin{subfigure}[t]{0.32\linewidth}
    \centering\hspace*{-0.5cm}
    \includegraphics[width=\linewidth]{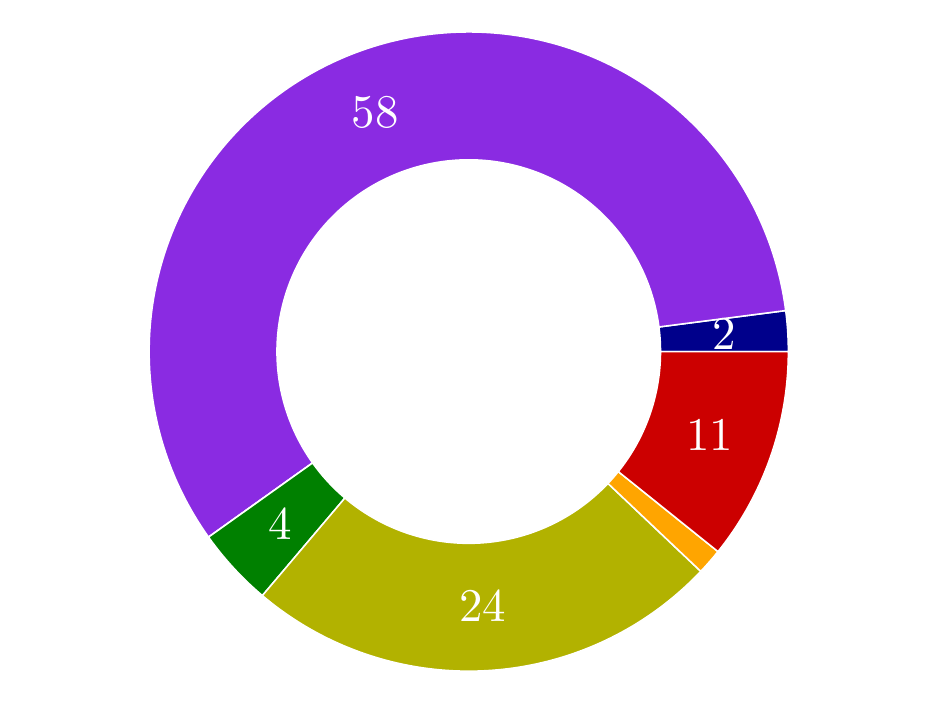}
    \caption{Sphere\hspace*{2em}}
  \end{subfigure}
  \begin{subfigure}[t]{0.325\linewidth}
    \centering
    \subcaptionbox{Moving cylinder\hspace*{5em}}{%
      \includegraphics[width=\linewidth]{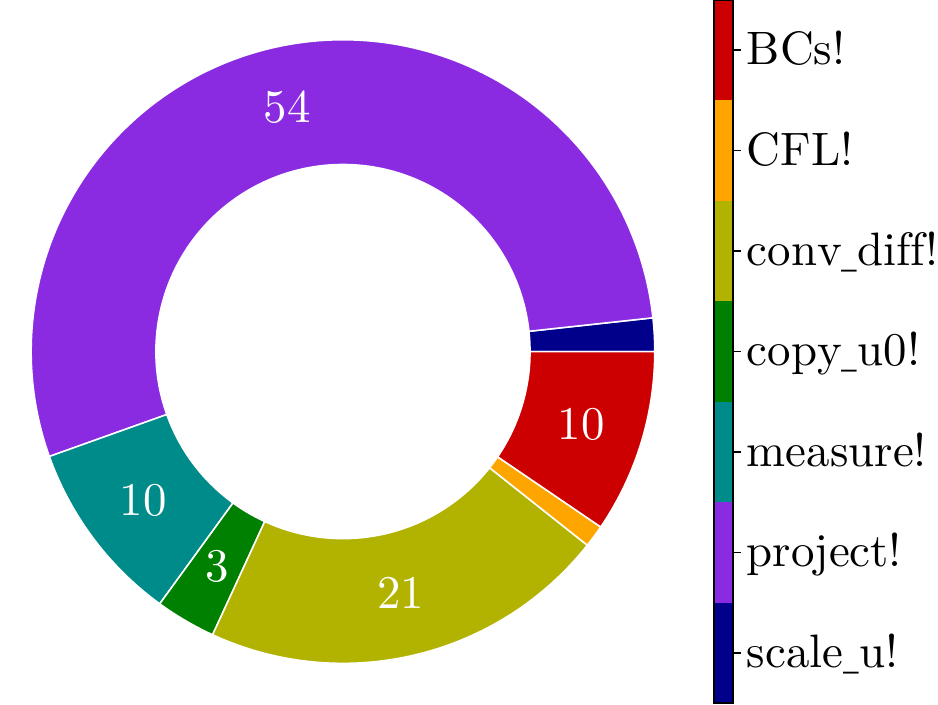}%
    }
  \end{subfigure}
  \caption{Execution time distribution (in percentage) of the routines in the time-stepping loop for the 3rd-level grids of the different test cases. Timings are measured as the median value of the routine execution time for 1000 time steps, noting that each routine can be called more than once for each time step (ie. predictor-corrector scheme). The following convention applies; \jlinl{scale_u!}: scalar operation that scales the velocity field; \jlinl{project!}: pressure-Poisson equation solver; \jlinl{measure!}: coordinates mapping for a moving solid boundary; \jlinl{conv_diff!}: computation of convective and diffusive terms; \jlinl{CFL!}: time step prediction; \jlinl{BCs!}: aggregated boundary conditions routines including: \jlinl{BC!}, \jlinl{BDIM!}, and \jlinl{exitBC!}, where the latter implements a convective outlet, and it is the most expensive boundary-condition routine.}
\label{fig:profiling}
\end{figure}

\begin{figure}[!t]
  \centering
  \begin{subfigure}[t]{\linewidth}
    \centering
    \includegraphics[width=0.65\linewidth]{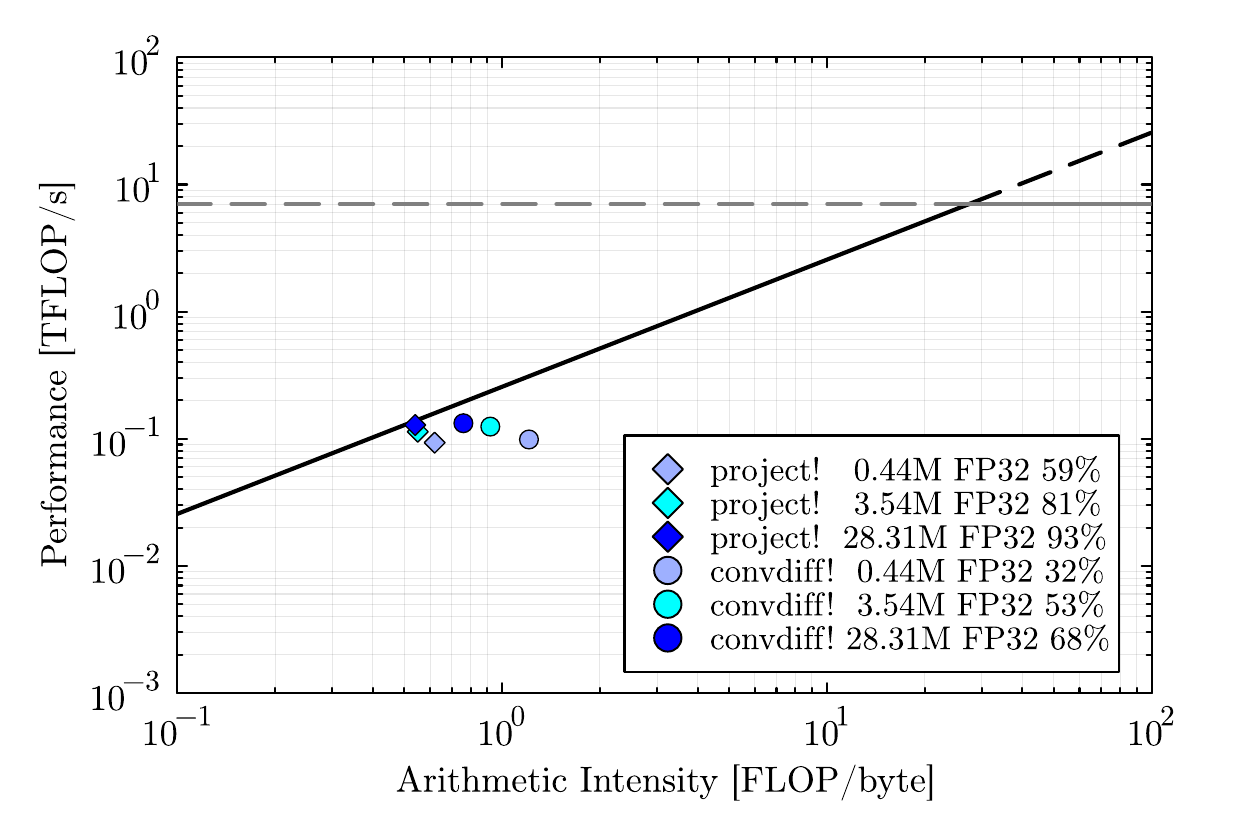}
    \caption{Roofline model obtained with Nsight Compute displaying the performance of the most expensive kernel in the convection/diffusion routine (kernel 355) and the pressure solver (kernel 327). The percentage value indicates the measured bandwidth (Performance/Arithmetic Intensity) for each kernel with respect to the maximum memory bandwidth (black line, 256 GB/s). Grey line: FP32 peak performance.}
    \label{fig:roofline}
    \vspace*{1em}
  \end{subfigure}
  \begin{subfigure}[t]{\linewidth}
    \centering
    \includegraphics[width=\linewidth]{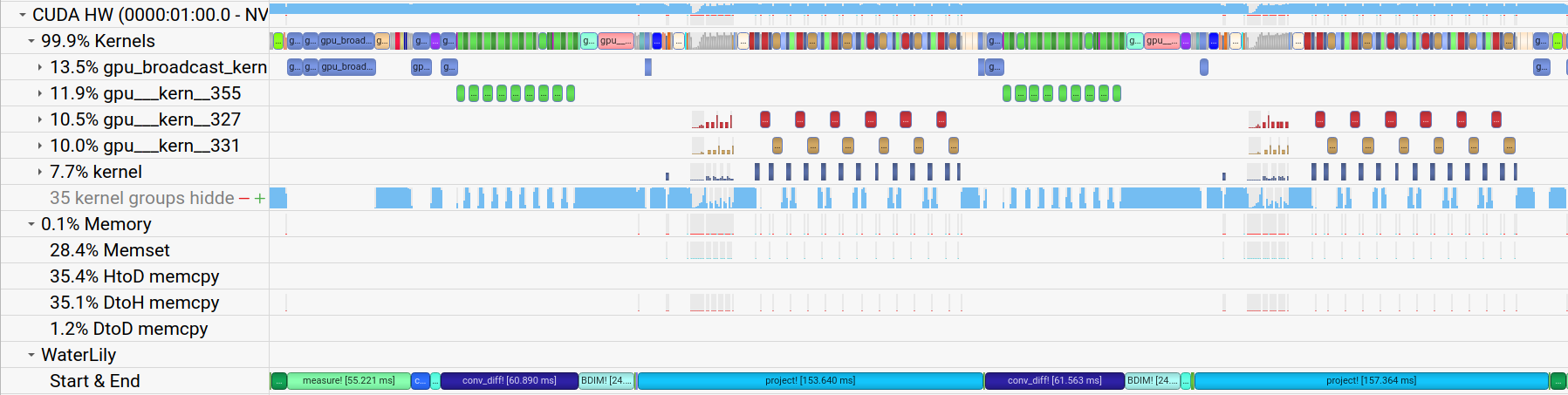}
    \caption{Nsight Systems analysis of the kernels used in one time step. The distribution of the NVTX ranges is displayed in the bottom row.}\label{fig:nsys_screenshot}
  \end{subfigure}
  \caption{Kernels profiling using NVIDIA Nsight Systems and Compute on the 3rd-level grid of moving cylinder test case.}
\label{fig:profiling_nsight}
\end{figure}

The profiling of the solver is conducted on a NVIDIA GeForce RTX 4060 laptop GPU using FP32 simulations by timing the execution range of the routines within the time-stepping loop using the NVTX.jl profiling package \citep{Byrne2023} (a wrapper of the NVIDIA Tools Extension Library, NVTX). We note that, as mentioned in \sref{sec:software_design}, kernels are automatically generated using macro expressions and the resulting kernel names are arbitrary. Hence, an NVTX range which traces a particular routine in the solver may contain more than one kernel, as displayed in \fref{fig:nsys_screenshot}.

The execution time distribution of the main routines in the time-stepping loop is displayed in \fref{fig:profiling} for the different cases. Noticeably, the \jlinl{project!} routine, ie. the pressure solver routine, and the \jlinl{conv_diff!} routine, ie. the convection and diffusion routine, consume most of the execution time for all cases. In addition, the cost of the \jlinl{measure!} routine in the moving cylinder case, which is used to map the solid boundary to a new position at every time step, accounts for 10\% of the time step cost. Regarding the pressure solver, it is worth noting that a preconditioned conjugate gradient smoother (PCG) is employed in the geometric multi-grid solver. Computing dot products (array reductions) on a GPU is known to be rather inefficient because of their low arithmetic intensity, and the PCG solver contains several reduction operations. The arithmetic intensity of the main kernels within \jlinl{project!} and \jlinl{conv_diff!} is displayed in the roofline model of \fref{fig:roofline}. It is shown that the arithmetic intensity of the dominant kernel in \jlinl{project!} is indeed lower that the one in \jlinl{conv_diff!} for a given grid size.

Moreover, \fref{fig:roofline} demonstrates that the main kernels, and the overall solver by extension, are bounded by the maximum memory bandwidth of the GPU. This is expected for a CFD solver which typically performs a reduced number of operations on the data loaded into the computing cores, hence yielding low arithmetic intensity.
Using the Nsight Compute tool to assess the GPU pipes utilization shows that main convection/diffusion kernel and the main kernel in the pressure solver mostly use the arithmetic logic unit (ALU) pipe (60\% of active cycles, approximately), and the fused multiply add/accumulate (FMA) pipe (40\% of active cycles, approximately). The measured pipe utilization is similar for all grids, but slightly increasing with grid size.

The Nsight Compute tool is also used to analyse the memory workload of the kernels. It is reported that data is only accessed from the global memory pool of the L1 cache. Loading from global memory is an additional bottleneck which could be improved by using the shared memory pool instead. With this, future work to improve the overall performance will focus on designing a better memory mapping which maximizes the efficient use of memory resources. Additionally, the measured hit rates of the L1 cache are within 30\% to 50\%, and cache re-use should be implemented to increase performance as well. With respect to the data transfer between host (CPU) and device (GPU), the Nsight Systems profiling (\fref{fig:nsys_screenshot}) shows that 99.9\% of the time is spent in computing, while only 0.1\% is spent in memory-copy (\jlinl{memcpy}) operations. Small data transfer operations are detected in the pressure solver, which are automatically handled by KA, even though these are negligible compared to the measured compute workload. Also, we note that the allocation of the necessary flow fields and buffer arrays in the GPU vRAM is a one-time operation instructed at the start of a simulation, so it does not impact the performance of the solver. In terms of memory footprint, the finest cylinder simulation (226 million DOF) consumes 29 GB of memory using FP32.

\section{Validation} \label{sec:validation}
\begin{figure}[!t]
    \centering
    \includegraphics[width=0.7\linewidth]{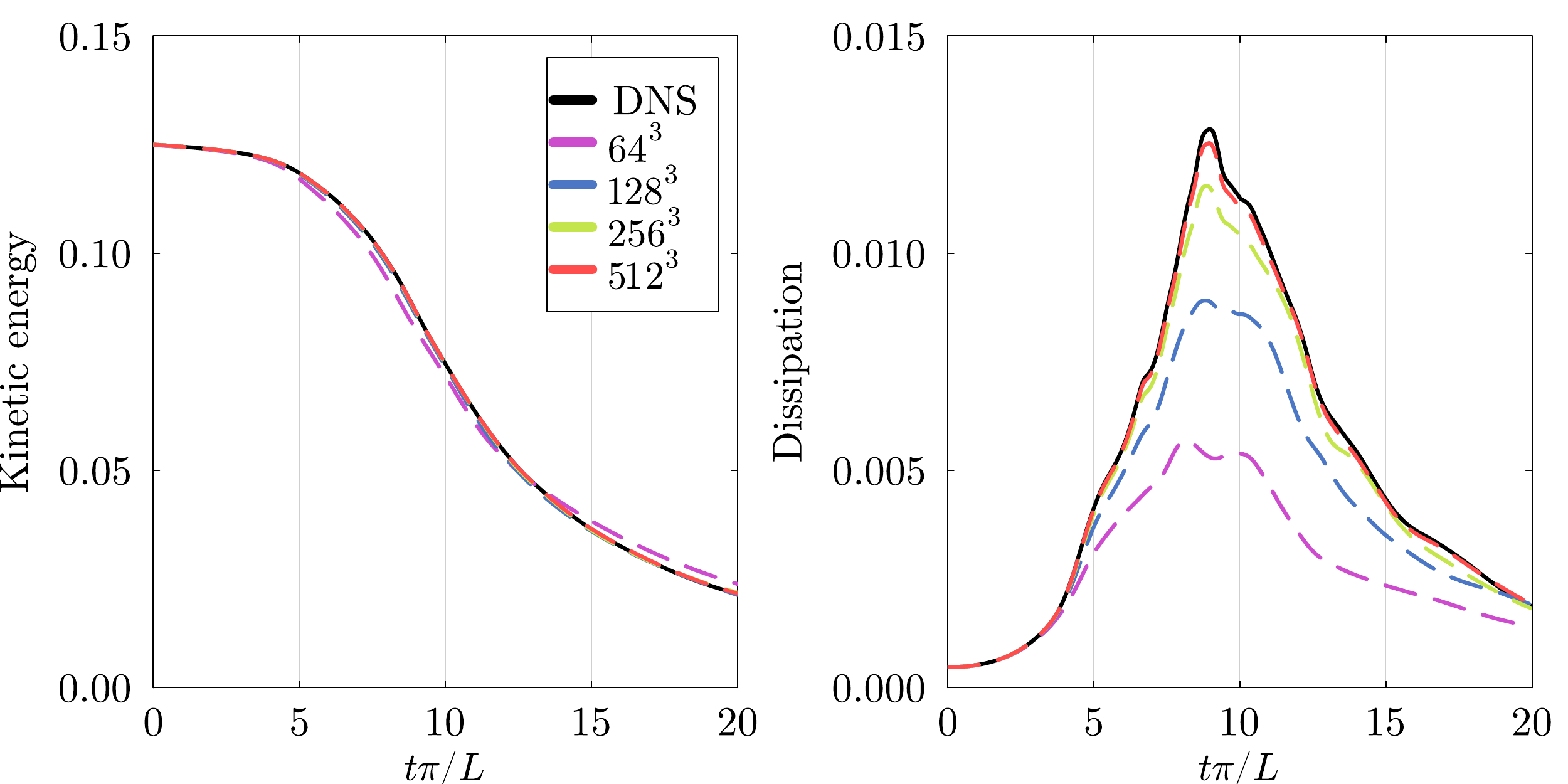}
	\caption{Taylor--Green vortex (TGV) temporal evolution of kinetic energy (left) and enstrophy (right). Direct numerical simulation (DNS) data from \cite{Dairay2017} is used as reference.}
	\label{fig:tgv_val}
\end{figure}

The validation of the solver is performed for the TGV, sphere, and an oscillating cylinder test cases. A grid convergence of the TGV case based on temporal evolution of kinetic energy and enstrophy is displayed in \fref{fig:tgv_val}. The finest grid consists of $512^3$ cells spanning the symmetric subdomain ($1/8$th) of the triple periodic box, similarly to the direct numerical simulation (DNS) reference data by \cite{Dairay2017}. The fine grid results greatly match the DNS reference data in which this same resolution was used.

\begin{figure}[!t]
  \centering
  \includegraphics[width=0.4\linewidth]{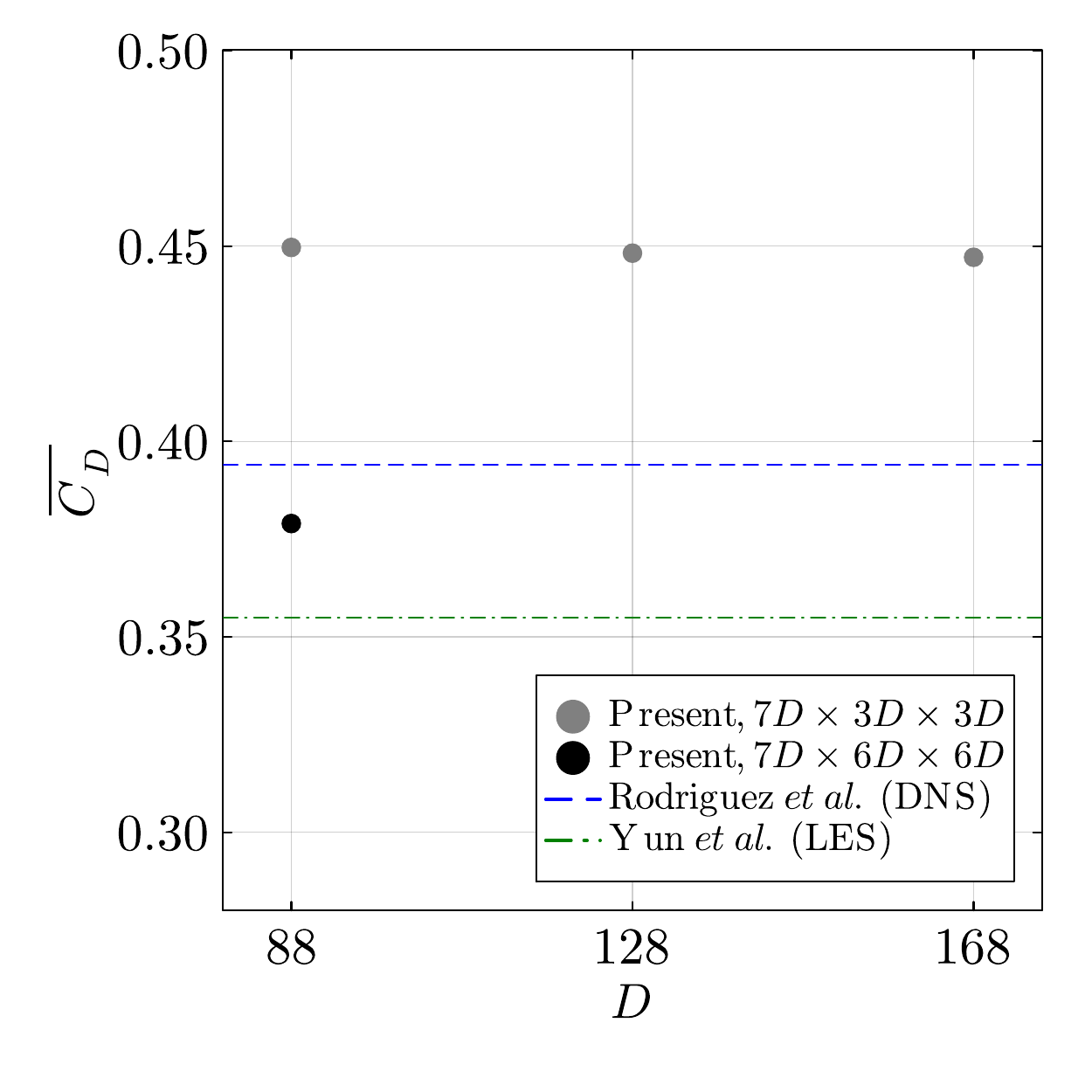}
  \vspace*{-0.5cm}
  \caption{Time-averaged drag coefficient measured on the sphere at $Re=3700$ for different resolutions (cells per diameter). The time-averaged metric is integrated over 300 convective time units (CTU, $tU/L$) after discarding the first 100 CTU used to reach the statistically-steady state of the wake.}
  \label{fig:sphere_val}
\end{figure}

For the sphere case, a domain of $7D\times 3D\times 3D$ and $7D\times 6D\times 6D$, where $D$ is the sphere diameter, are considered for validation. The reason for this is to allow for a greater resolution of the boundary layer, which would not be feasible with the domain used for benchmarking ($16D\times 6D\times 6D$). Two domain sizes are considered to assess the domain blockage. With this, the grids tested for validation contain 88, 128, and 168 cells
per diameter, totalling 43M, 132M, and 299M DOF on the small domain, respectively. The minimum boundary layer thickness around the sphere at $Re=3700$ is approximately $\delta /D=0.02$ \citep{Capuano2023}, and DNS studies such as \citet{Rodriguez2011} fit 12 grid points within the boundary layer thickness. In this case, the finest grid has a resolution of $h/D=0.006$ approximately, which means that only 3 grid points are used to represent the boundary layer. Hence, it is important noticing that the current Cartesian-mesh method using constant spacing requires vast resources to fully resolve the boundary layer of the immersed bodies. As shown in \fref{fig:sphere_val}, a converged time-averaged drag coefficient of approximately $\overline{C_{D}}=0.45$ is found for all the tested grids of the small domain case, and a drag coefficient of $\overline{C_{D}}=0.38$ is found for the large domain. These results show appropriate convergence in terms of grid resolution and domain size, and the large-domain case is within 4\% error of the DNS data from \citet{Rodriguez2011} and 7\% error of the LES data from \citet{Yun2006}.

\begin{figure}[!t]
    \begin{subfigure}{0.5\linewidth}
      \centering
        \includegraphics[width=1\linewidth,trim={130 70 20 50},clip]{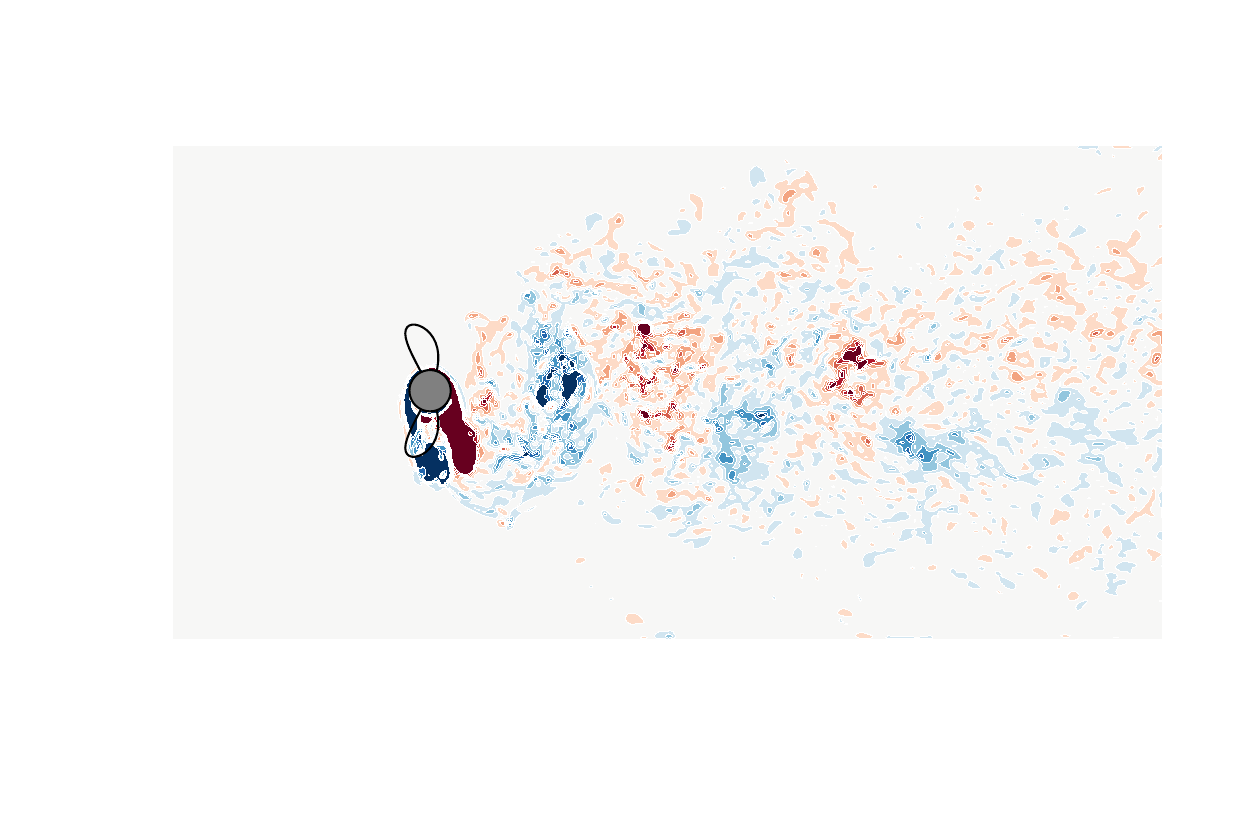}
        \caption{Cylinder path and spanwise-averaged vorticity field}
    \end{subfigure}
    \begin{subfigure}{0.48\linewidth}
        \centering
        \includegraphics[width=0.9\linewidth,trim={0 0 0 20},clip]{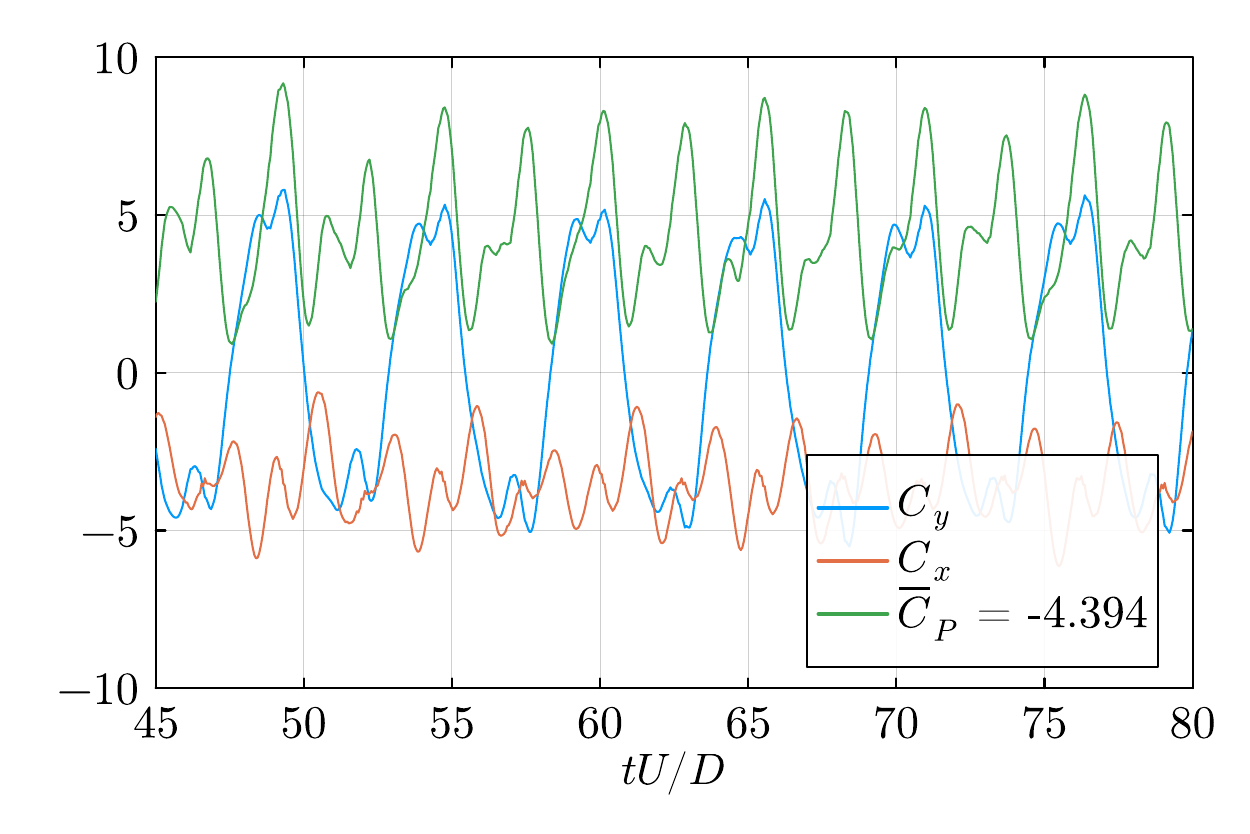}
        \caption{Force and power coefficients over six cycles}
    \end{subfigure}
  \caption{Oscillating cylinder validation. The wake is highly three-dimensional and turbulent, but spanwise-averaging helps visualize the main vortex structures and the integrated forces are still smooth.}
  \label{fig:cyl_val}
\end{figure}

To validate the moving immersed boundary method, we compare against the oscillating cylinder experiments of \citep{Aktosun2024}, with both cross-stream and inline motions:
\begin{equation*}
    y = A_y\sin(\omega t), \quad x = A_x\sin(2\omega t+\theta)
\end{equation*}
and we pick a large amplitude cases where $A_y=1.6D$, $A_x=0.4D$, $\theta=\pi/6$ and $\omega=\frac{2\pi}{5.4} U/D$ where $U,D$ are the inflow velocity and cylinder diameter, respectively, and the Reynolds number is $Re_D=7620$. The numerical domain is set to $4n \times 2n \times n$, where $n=2^7$ cells, giving 16.8M cells overall. Symmetry conditions are applied on all the domain boundaries other than the exit, which uses a convective outflow condition. The cylinder spans the full $z$ domain, is placed $n$ cells away from the side and inlet boundaries, and has a diameter of $D=n/6$. Figure \ref{fig:cyl_val}a shows the cylinder path and resulting vorticity field showing an extremely energetic 3D wake.  Figure \ref{fig:cyl_val}b shows the measured force and power coefficients, and the mean power coefficient of $\overline{C_P}=-4.39$ is within 3\% of the experimentally measured mean of -$4.52$ for this case.

\section{Sample applications}\label{sec:applications}
Three applications are selected to demonstrate the capability of the package to analyze general fluid flows. The examples also showcase the advantages of a differentiable backend-agnostic Cartesian-grid solver.

\subsection{Optimized control cylinders}

\begin{figure}
    \centering
    \begin{subfigure}[t]{0.8\linewidth}
        \centering
        \includegraphics[width=\linewidth,trim={50 70 20 210},clip]{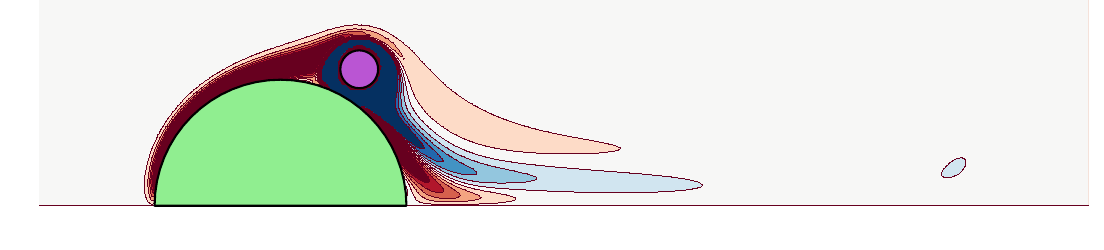}
        \vspace{-1cm}
        \caption{Vorticity field}
    \end{subfigure}
    \begin{subfigure}[b]{0.38\linewidth}
        \includegraphics[width=\linewidth]{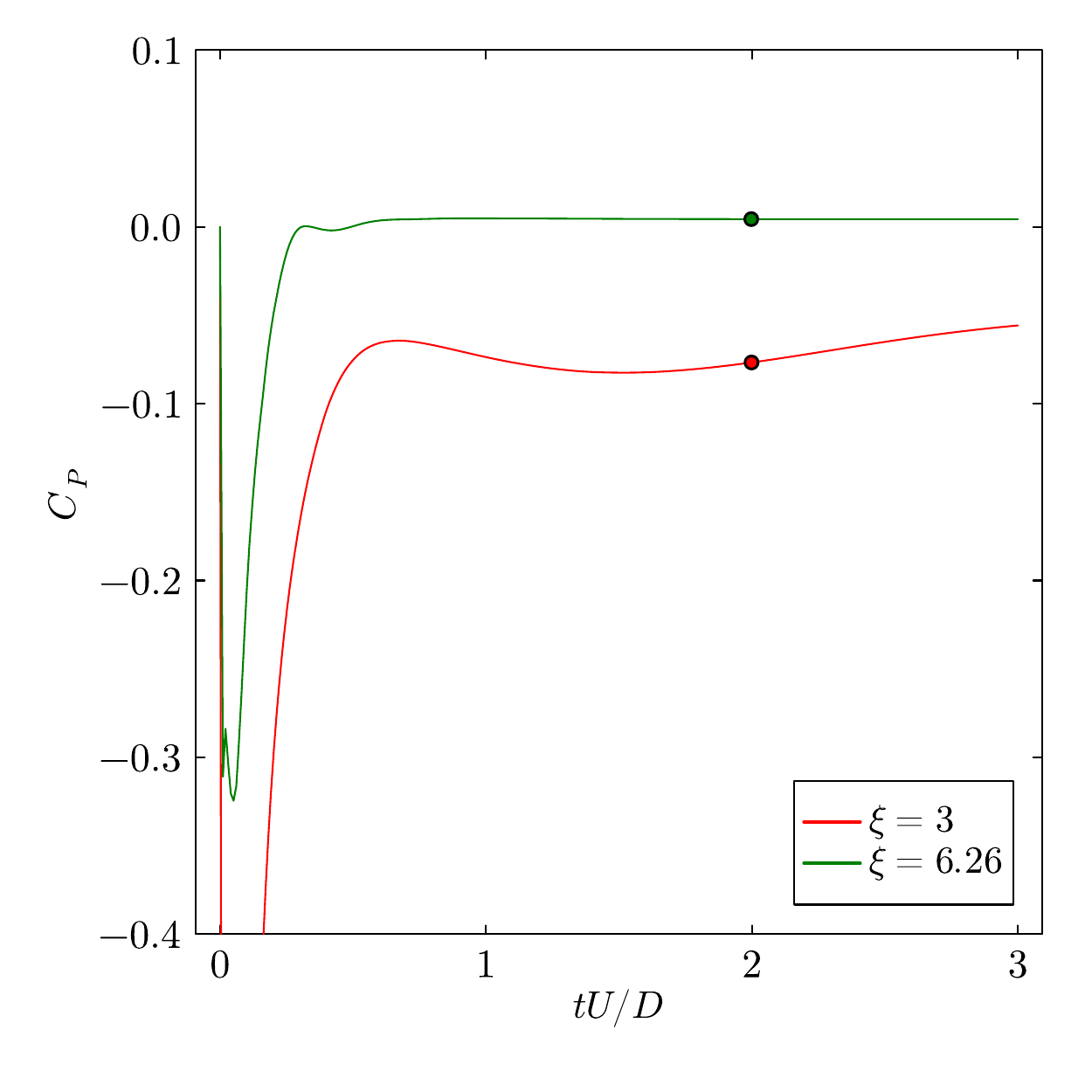}
        \vspace{-1cm}
        \caption{Scaled power history}
    \end{subfigure}\hspace{20pt}
    \begin{subfigure}[b]{0.38\linewidth}
        \centering
        \includegraphics[width=\linewidth]{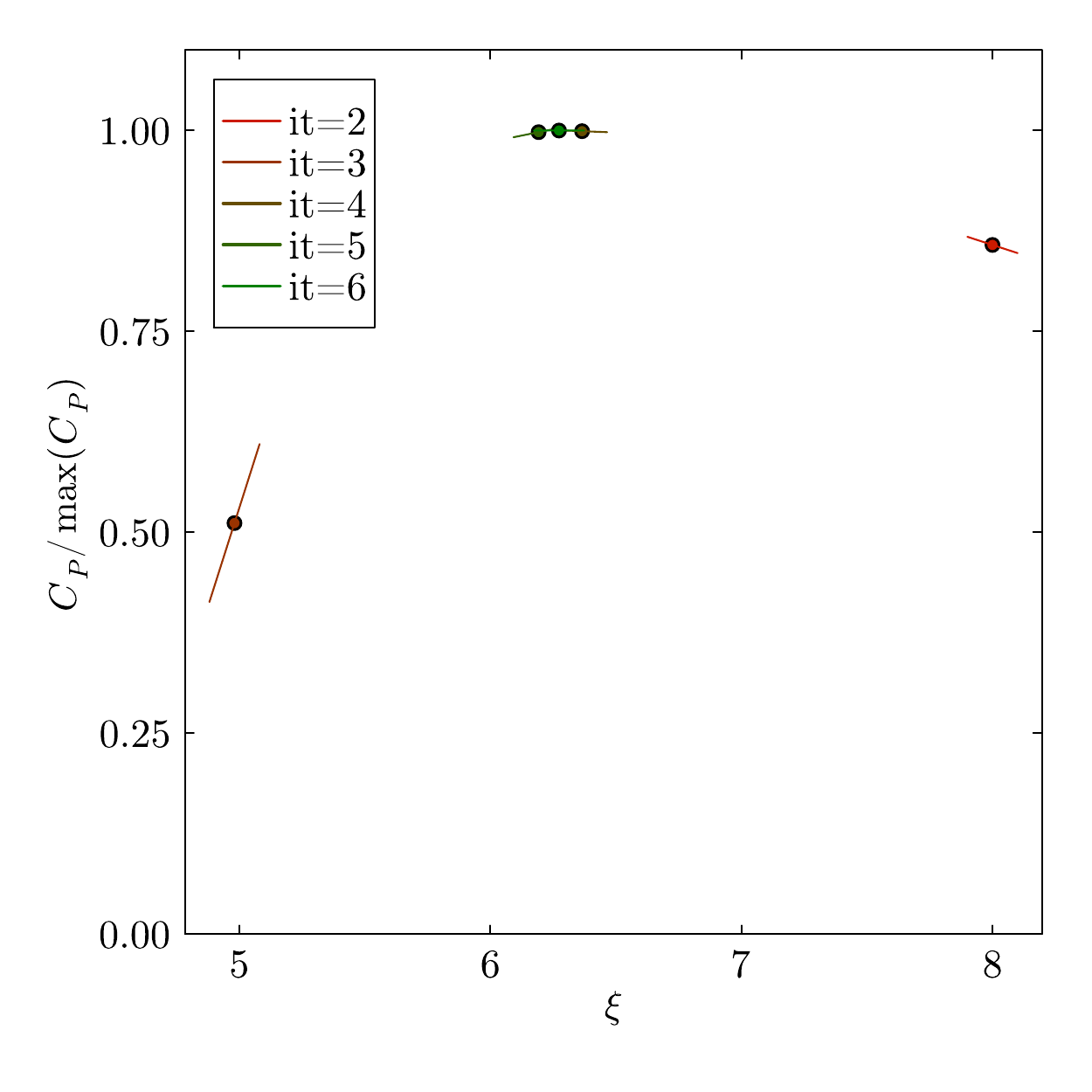}
        \vspace{-1cm}
        \caption{Optimization process}
    \end{subfigure}
    \caption{Controlled flow over a static cylinder using spinning cylinders in the wake. The small cylinder (purple) has scale spin velocity $\xi$, driving the flow to be symmetry and steady after an initial transient, as measured by the vorticity (a) and the scaled power coefficient $C_P$ (b). The scaled $C_P$ (points) and its derivative (sloped lines) are used to determine the optimum spin in 6 iterations (c).}
    \label{fig:spinning_circle}
\end{figure}

The first example will be optimizing the controlled 2D flow around a circle using a pair of small spinning circles placed 120 degrees relative to the inflow direction, \fref{fig:spinning_circle}a. Experimental and numerical studies of this system have shown the capability of the spinning cylinders to control the flow over the large circle \citep{Schulmeister2017}, establishing a steady symmetric wake, reducing the system drag and even producing a net thrust as the rotation rate is increased.

The system is described by a few dimensionless ratios: $Re=UD/\nu$ the Reynolds number based on the large circle diameter and inflow velocity, $d/D$ the scaled diameter of the control circle, $g/D$ the gap between the large circle and the control circle, and $\xi=\frac 12 d\Omega/U$ the control circle scaled surface speed. This system is simulated with WaterLily using the values of $Re=500,\ d/D=0.15,\ g/D=0.05$ with grid resolution $D/h=96$. The domain is sized to $6D\times2D$ taking advantage of the known symmetry of the flow by using a symmetry plane and only modelling the upper half of the full domain. The entire differentiable simulation is defined with the simple script

\begin{minipage}{\linewidth}\noindent
\begin{jllisting}
rot(θ) = [cos(θ) -sin(θ); sin(θ) cos(θ)] # rotation matrix
function drag_control_sim(ξ; D=96, Re=500, d_D=0.15f0, g_D=0.05f0)
    # set up big cylinder
    C, R, U = [2D, 0], D÷2, 1
    big = AutoBody((x, t) -> √sum(abs2, x - C) - R) # signed-distance function

    # set up small control cylinder
    r = d_D * R
    c = C + (R + r + g_D * D) * [1 / 2, √3 / 2]
    small = AutoBody(
        (x, t) -> √sum(abs2, x) - r,           # signed-distance function
        (x, t) -> rot(ξ * U * t / r) * (x - c) # center and spin!
    )

    # set up simulation
    Simulation((6D, 2D), (U, 0), D; ν=U * D / Re, body=big + small, T=typeof(ξ))
end
\end{jllisting}
\end{minipage}

This example demonstrates that WaterLily can combine \jlinl{AutoBody} types based on the arithmetic of signed-distance functions. The two \jlinl{big} and \jlinl{small} circles are defined with a line of code each and combined trivially with \jlinl{body = big + small}. It also demonstrates that the variable $\xi$ is used to set the types employed for the simulation. This allows easy switching between any floating point precision, and enables AD to be applied to the solver as a whole by running the code with a \jlinl{T = Dual} data-type holding the value and derivative simultaneously \citep{RevelsLubinPapamarkou2016}. As discussed in the software design section above and in that reference, this generates efficient code for the function and the exact derivative (not a finite-difference approximation), at a cost only 80\% larger than evaluating the function alone.

We use the differentiable solver to maximize the scaled propulsive power $C_P = FU/\rho dc^3$ where $F$ is the net thrust force on the system. This metric is  proportional to the propulsive efficiency since $\rho dc^3$ scales with the power required to rotate the control cylinders \citep{Schulmeister2017}. The time history of $C_P$ is plotted for two values of $\xi$ in \fref{fig:spinning_circle}b, demonstrating that only a few convective cycles are required to reach steady state, as well as the control authority of $\xi$ over the propulsive power.

We optimize $\hat\xi=\text{argmax}\ C_P(\xi)$ at time $t^*=tU/L=2$ using Davidon's method \citep{Davidon1991}, which evaluates $C_P$ and its derivative $\partial C_P/\partial \xi$ at points bracketing an optimum, using inverse cubic interpolation to iteratively restrict the interval. \fref{fig:spinning_circle}c shows the values of $C_P$ and its derivative $\partial C_P/\partial \xi$ at each value of $\xi$ during the optimization process, starting with the interval $\xi=[3,8]$, and leading to the optimum $\hat\xi\approx 6.26$ in a few iterations. Rates above this optimum produce more net thrust, but require excessive power to produce.

\subsection{Deforming and dynamic geometries}
\begin{figure}
    \centering
    \includegraphics[width=0.24\linewidth,trim={90 150 70 200},clip]{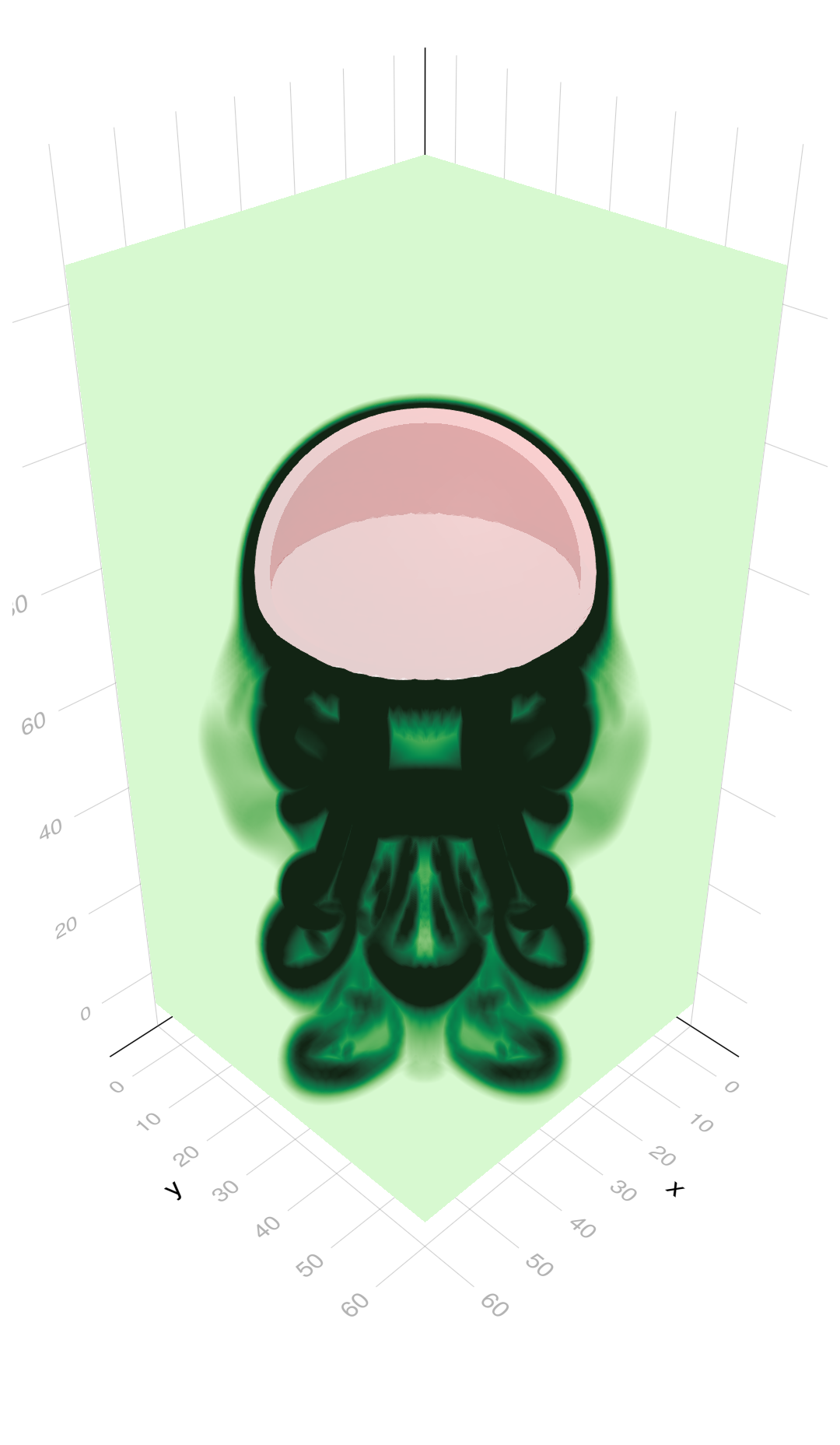}
    \includegraphics[width=0.24\linewidth,trim={90 150 70 200},clip]{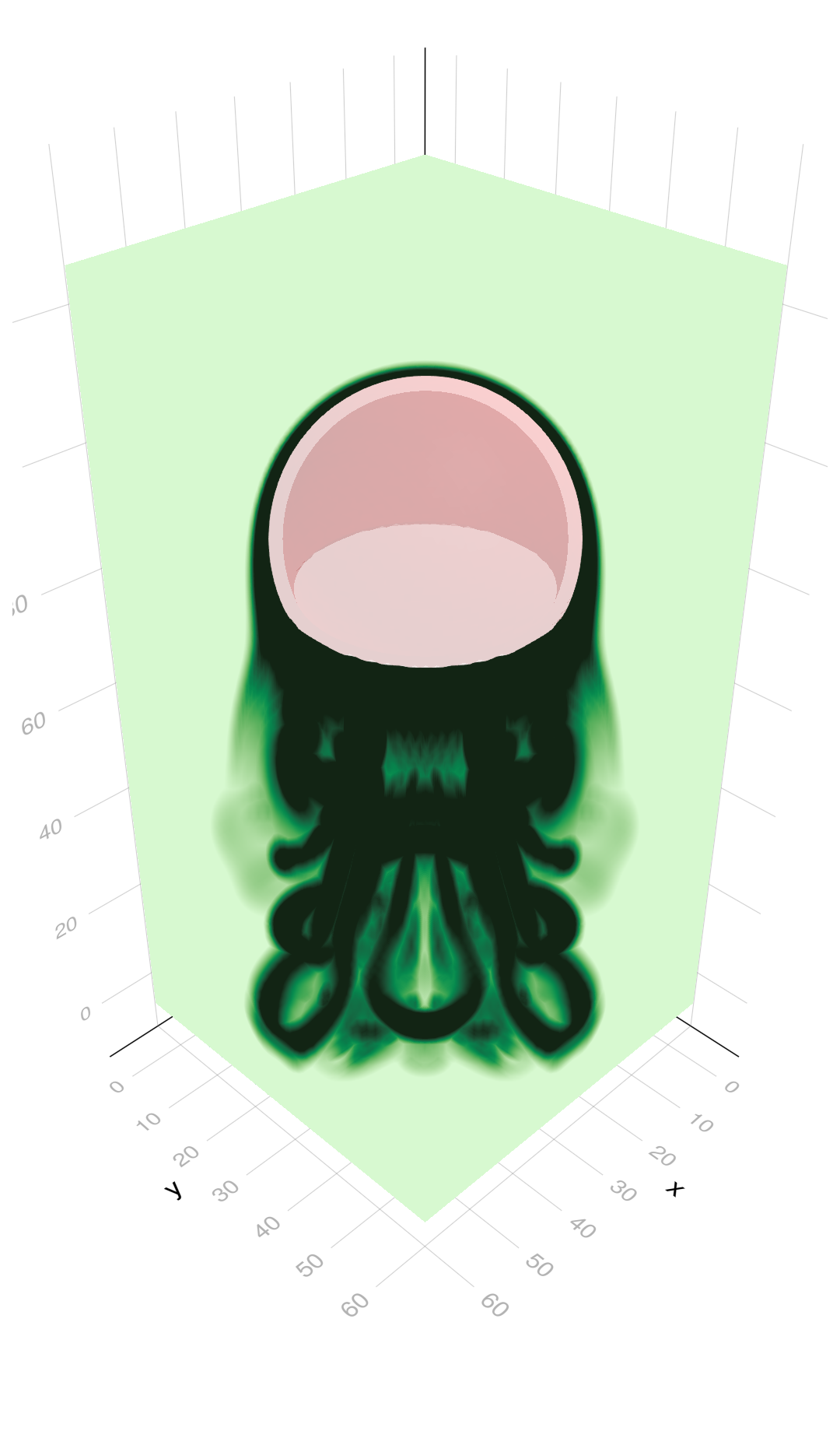}
    \includegraphics[width=0.24\linewidth,trim={90 150 70 200},clip]{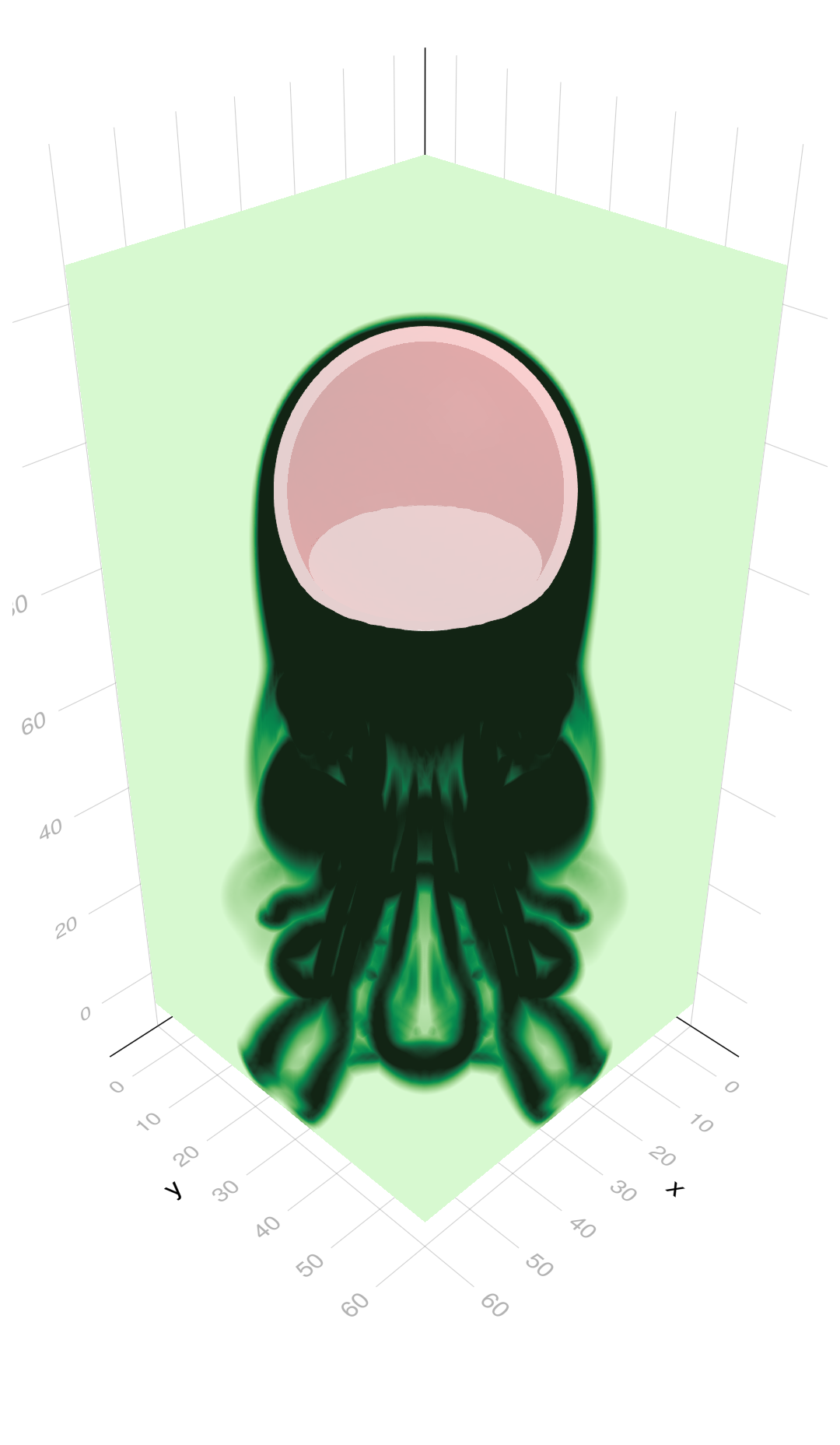}
    \includegraphics[width=0.24\linewidth,trim={90 150 70 200},clip]{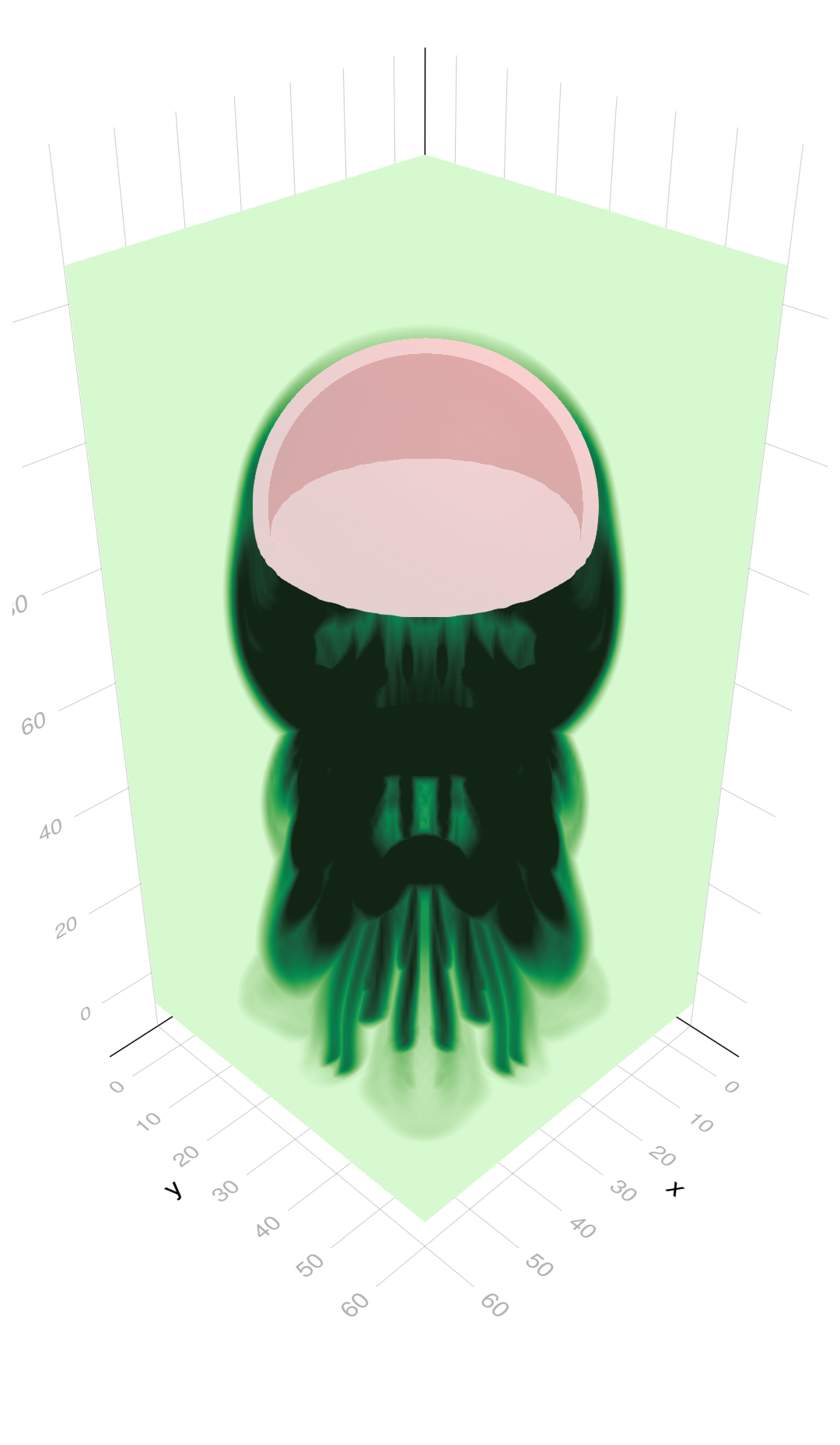}
    \caption{Flow induced by a pulsing jellyfish geometry visualized by vorticity magnitude at equally spaced intervals over a cycle.}
    \label{fig:jelly}
\end{figure}

The final two examples showcase the solver's ability to handle more complex geometries with ease. The first is a pulsing jellyfish-inspired geometry, and the second is a whale tail-inspired geometry. Both of these cases are fast enough to simulate on a laptop GPU for live demonstrations at reduced resolution.

\begin{minipage}{\linewidth}
\begin{jllisting}
function jelly(p=5; Re=500, mem=CuArray, U=1)
    # Define simulation size, geometry dimensions, & viscosity
    n = 2^p
    R = 2n / 3
    h = 4n - 2R
    ν = U * R / Re

    # Motion functions
    ω = 2U / R
    A(t) = 1 .- [1, 1, 0] * 0.1 * cos(ω * t)
    B(t) = [0, 0, 1] * ((cos(ω * t) - 1) * R / 4 - h)
    C(t) = [0, 0, 1] * sin(ω * t) * R / 4

    # Build jelly from a mapped sphere and plane
    sphere = AutoBody(
      (x, t) -> abs(√sum(abs2, x) - R) - 1, # sdf
      (x, t) -> A(t) .* x + B(t) + C(t)     # map
    )
    plane = AutoBody((x, t) -> x[3] - h, (x, t) -> x + C(t))
    body =  sphere - plane

    # Return initialized simulation
    Simulation((n, n, 4n), (0, 0, -U), R; ν, body, mem, T=Float32)
end
\end{jllisting}
\end{minipage}

The bell of the jellyfish is constructed with more \jlinl{AutoBody}-arithmetic, in this case taking the difference of a hollow sphere with an oriented plane. This geometry is made to pulse by mapping the coordinates harmonically in the radial and transverse directions. While the geometry maintains a roughly constant solid volume throughout the pulse, small deviations are handled gracefully by the solver. \fref{fig:jelly} shows equally spaced snapshots of the geometry and resulting flow throughout the cycle. Each cycle generates a strong propulsive vortex ring which breaks up as it propagates away, in qualitative agreement with experimental studies such as \cite{Dabiri2005}.

\begin{figure}
    \centering
    \includegraphics[width=0.32\linewidth,trim={550 250 530 400},clip]{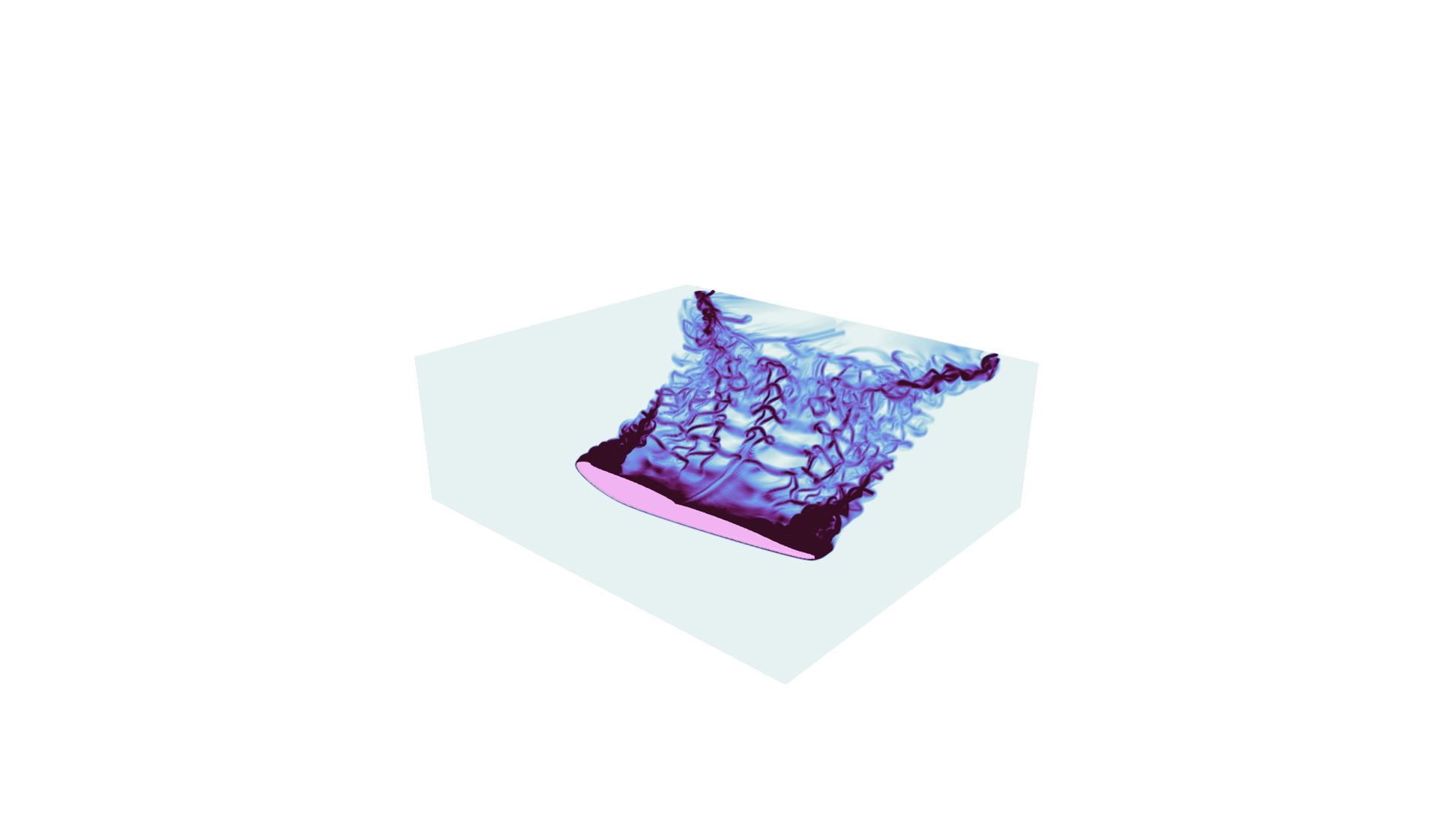}
    \includegraphics[width=0.32\linewidth,trim={550 250 530 400},clip]{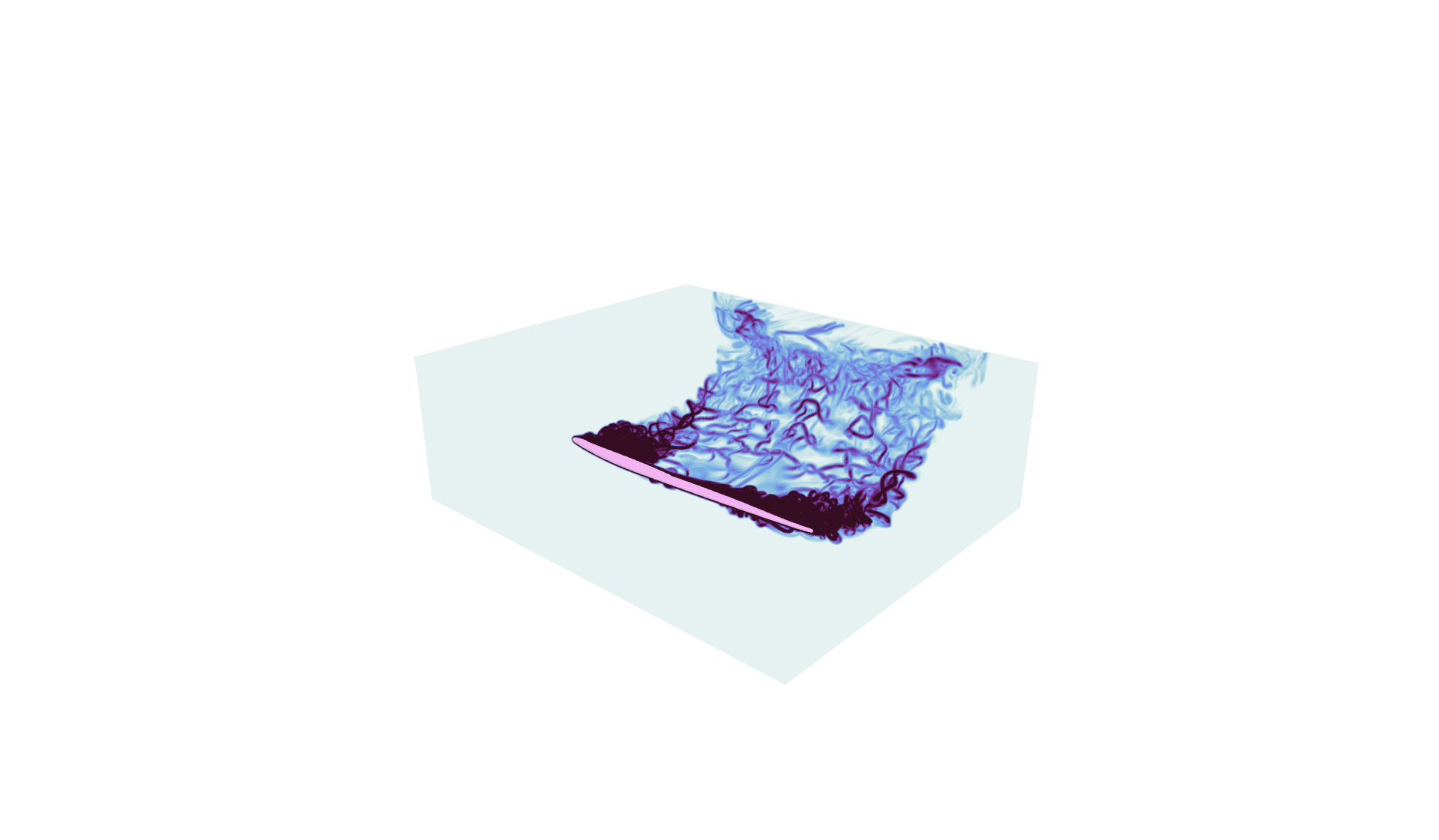}
    \includegraphics[width=0.32\linewidth,trim={550 250 530 400},clip]{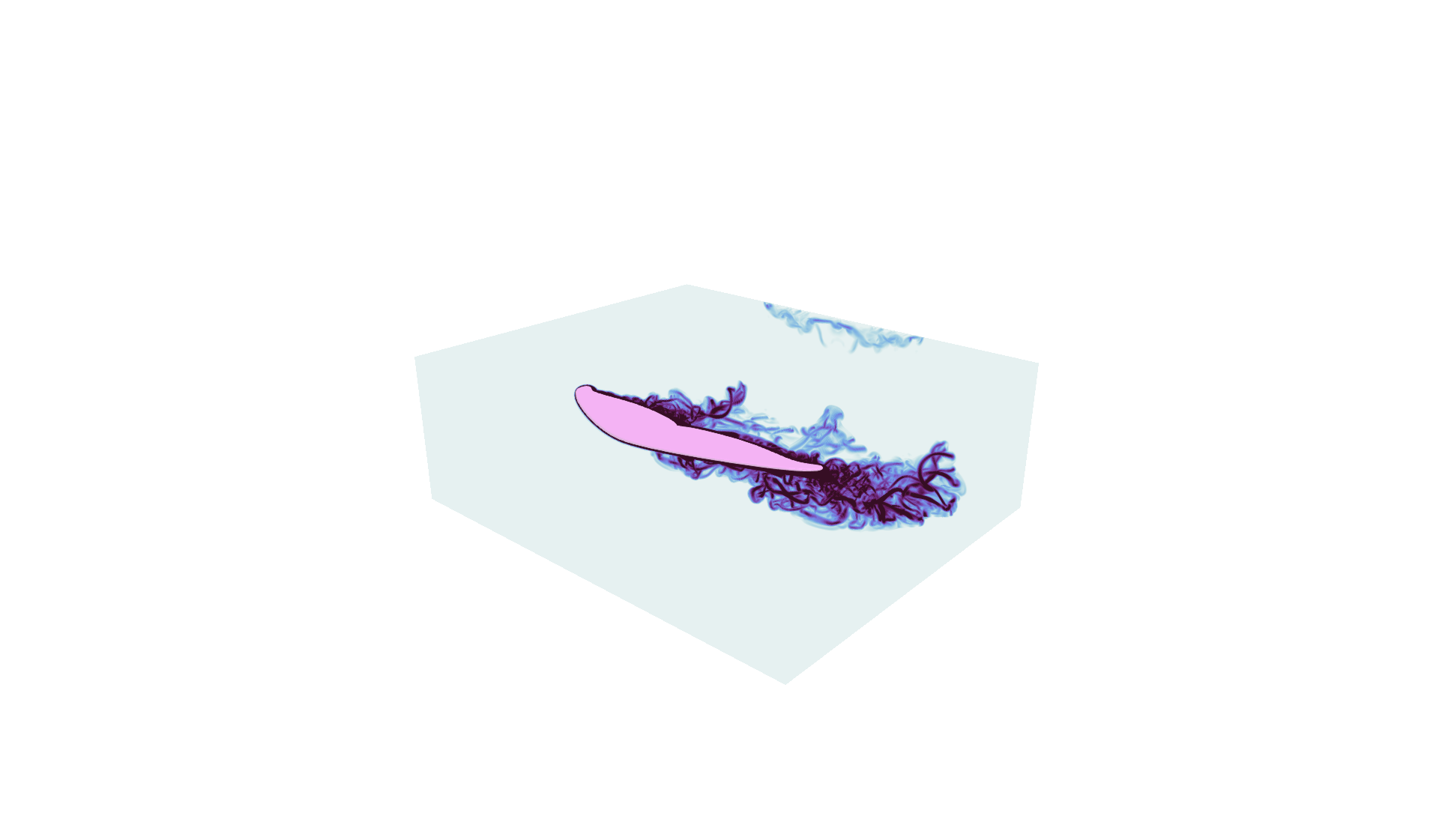}
    \includegraphics[width=0.32\linewidth,trim={550 250 530 400},clip]{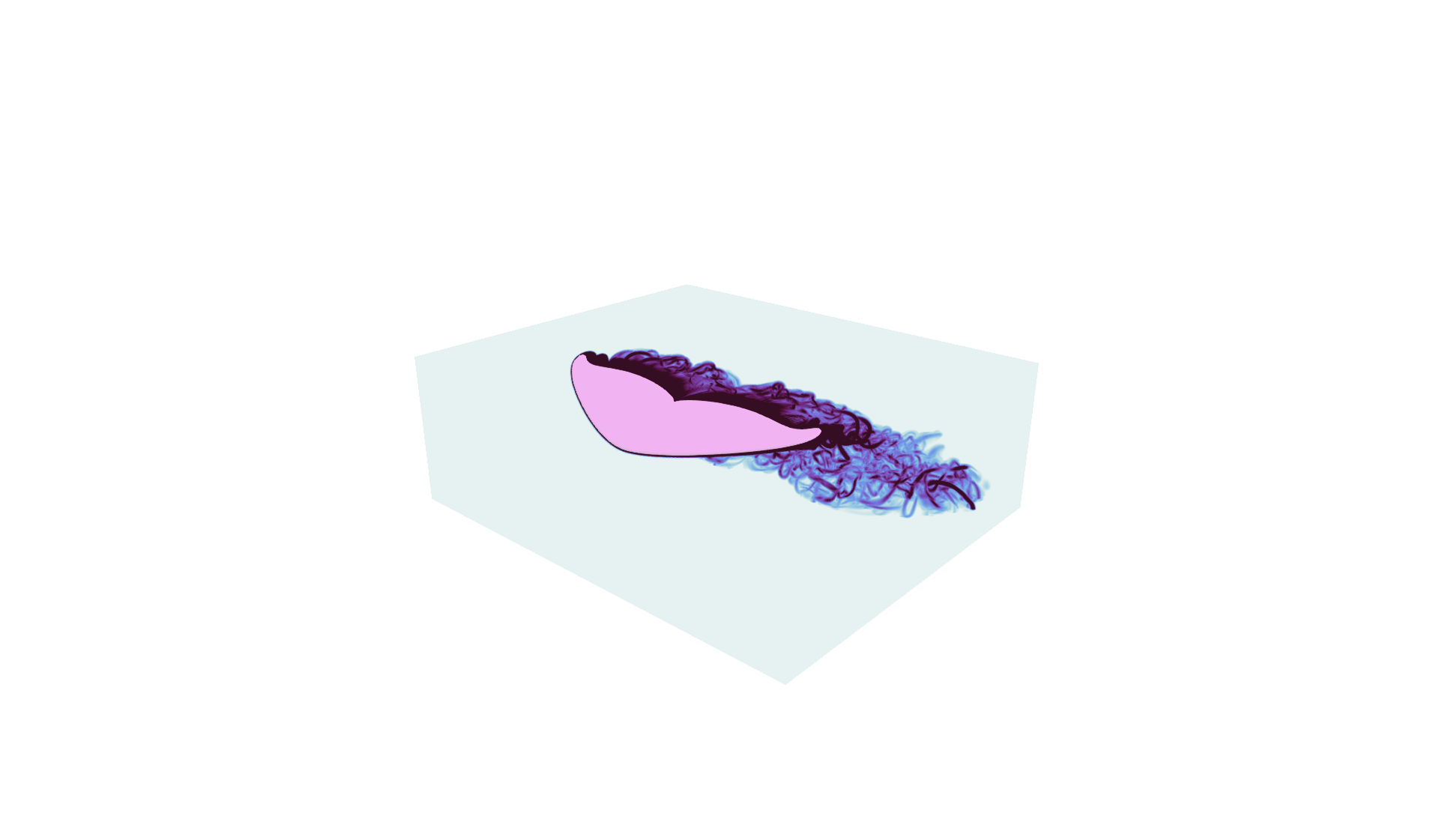}
    \includegraphics[width=0.32\linewidth,trim={550 250 530 400},clip]{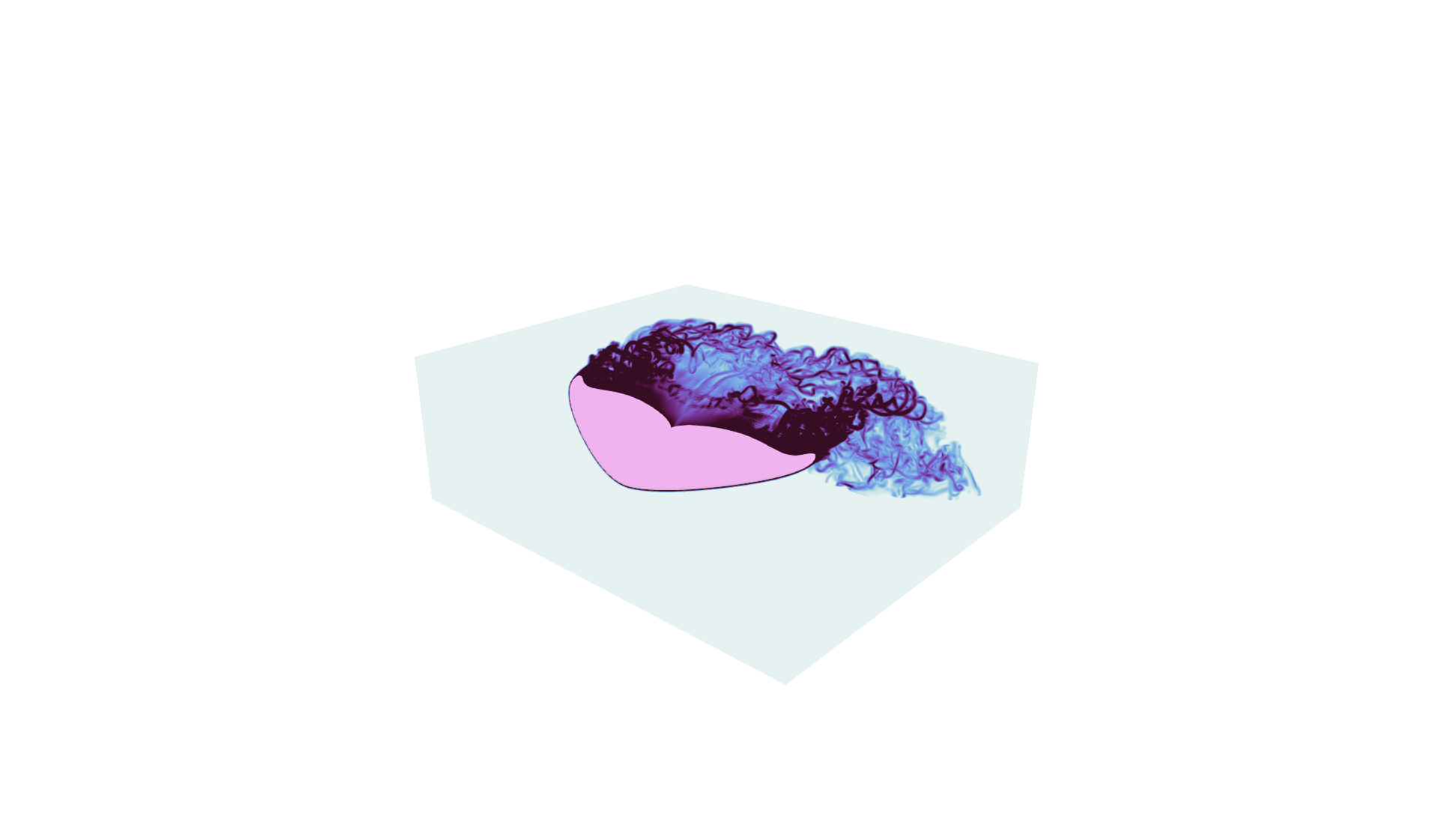}
    \includegraphics[width=0.32\linewidth,trim={550 250 530 400},clip]{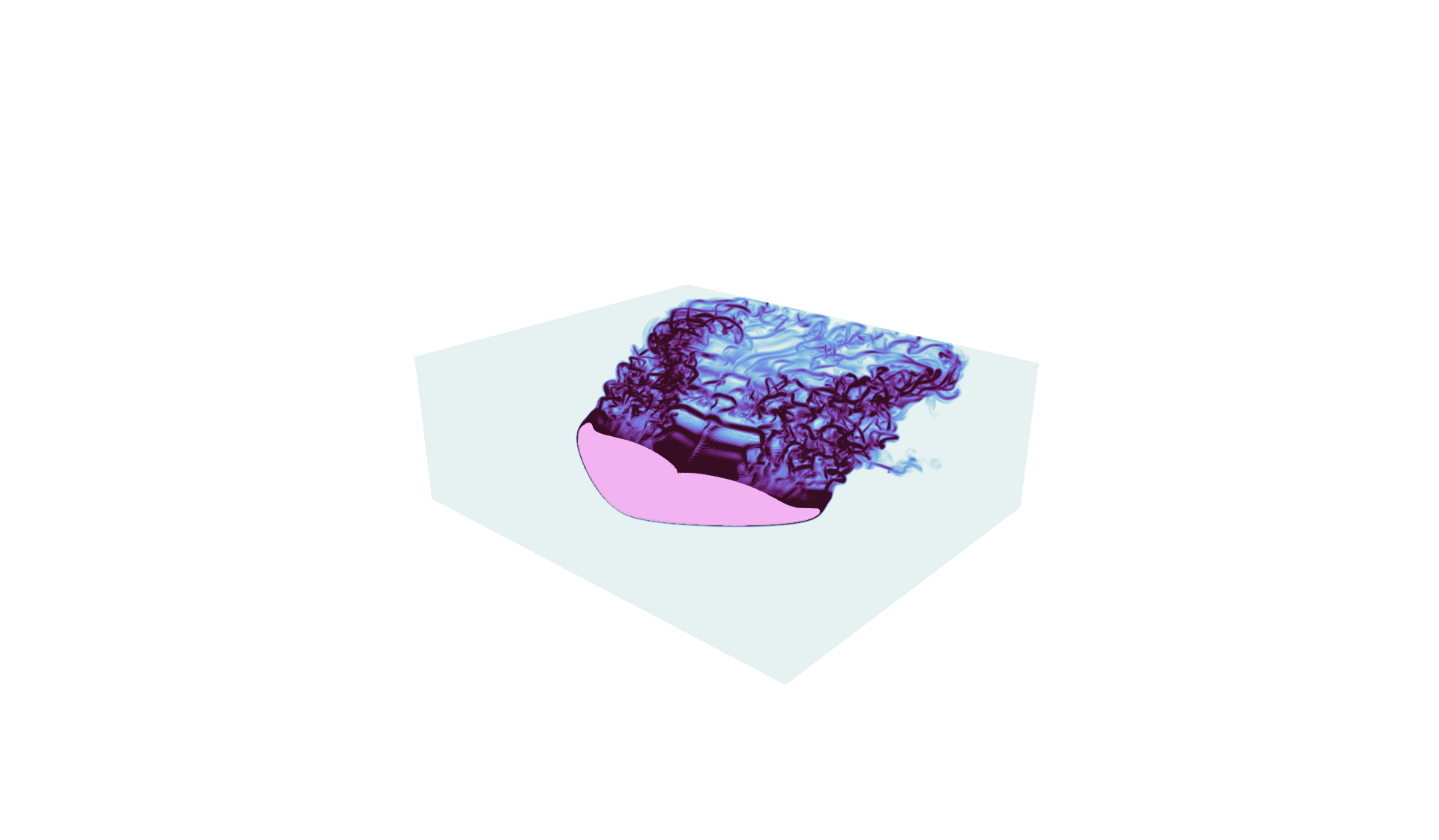}

    \caption{Flow induced by a flapping whale tail geometry visualized by vorticity magnitude at equally spaced intervals over a cycle using chord resolution $L=96$ and sweep $s=10$.}
    \label{fig:whale}
\end{figure}

\begin{minipage}{\linewidth}
\begin{jllisting}
function whale(s=10; L=24, U=1, Re=1e4, T=Float32, mem=CuArray)
    pnts = [
        0 40 190   200   190   170   100   0 -10 0
        0  0  8s 8s+40 5s+70 5s+50 5s+70 100  80 0
    ]
    planform = BSplineCurve(reverse(pnts), degree=3)
    function map(x, t)
        θ, h = π / 6 * sin(U * t / L), 1 + 0.5cos(U * t / L)
        Ry = SA[cos(θ) 0 sin(θ); 0 1 0; -sin(θ) 0 cos(θ)]
        Ry * 100 * (x / L - SA[0.75, 0, h])
    end
    body = PlanarParametricBody(planform, (0, 1); map, mem)
    Simulation((5L, 3L, 2L), (U, 0, 0), L; U, ν=U * L / Re, body, T, mem)
end
\end{jllisting}
\end{minipage}

The final example of the whale tail is a planar membrane defined using the \href{https://github.com/WaterLily-jl/ParametricBodies.jl}{ParametricBodies.jl} package \citep{WeymouthLauber2023}. The planform is defined by a set of points which are interpolated using a cubic spline. The sweep of the wing is adjustable with the input parameter $s$, allowing parametric geometry studies to be carried out with ease, as in \cite{ZurmanNasution2021}. The 3D distance function and normal to this planar membrane are then evaluated using a parametric root-finding method to immerse the geometry in the simulation as with the examples above. Harmonic pitch and heave motion are used to flap the tail. \fref{fig:whale} shows equally spaced snapshots of the geometry and resulting flow throughout the cycle, matching the vortex structures found in previous works \citep{ZurmanNasution2021}.

\section{Conclusions}\label{sec:conclusions}

In this work, an incompressible viscous flow solver written in the Julia language has been presented, namely WaterLily.jl. With a minimal codebase (less than 600 lines of code), WaterLily implements an $n$-dimensional CFD solver based on a Cartesian-grid finite-volume method which is able to handle arbitrary moving bodies through the boundary data immersion method. Using Julia's high-level libraries such as KernelAbstractions.jl and ForwardDiff.jl, the solver is able to run on diverse architectures (serial CPU, multithread CPU, and GPUs of different vendors), and it offers full differentiability based on automatic differentiation (AD).

Based on three different cases (TGV, fixed sphere, and moving cylinder), benchmarking results show that execution on a modern GPU can yield two orders of magnitude speed-up factors compared to serial CPU execution. Profiling on a GPU backend shows that the pressure solver and the convection-diffusion routines incur the highest cost during a time step. Additionally, performance analysis based on the roofline model shows that the solver is bounded by the memory bandwidth limit of the GPU. Furthermore, validation results demonstrate the accuracy of the solver on the three test cases. As the immersed boundary cases use AD within the solver to determine the normals and curvatures of the immersed geometry, these cases also validate the AD application.

In addition, we provide an example of using the differentiable solver to optimize the flow control with rotating cylinders. The derivative of the propulsive power coefficient is used to quickly optimize the spinning rate of the small cylinder controlling the wake of the large static cylinder for maximum thrust with minimum power. Finally, the possibility of simulating complex dynamic bodies is showed with a jellyfish-inspired geometry that heaves while also expanding and contracting, and a parametrically-defined flapping whale tail. Future work will focus on the parallelization of the solver at the distributed-memory level using the message passing interface (MPI) standard, the inclusion of multiphase flow simulation through the volume-of-fluid method, and the continuous improvement of the performance of the solver.

\section{Acknowledgements}\label{sec:acknowledgements}
The authors acknowledge the Barcelona Supercomputing Center for awarding access to the MareNostrum5 system. The authors also acknowledge Dr. Valentin Churavy for creating KernelAbstractions.jl and his continued support, and Dr. Lucas Gasparino for fruitful discussions and initial tests on his personal GPU.

\bibliographystyle{etc/apalike-refs}
\bibliography{main}

\end{document}